\documentclass[11pt,a4]{iopart}
\usepackage[utf8]{inputenc}

\usepackage{amsbsy}
\usepackage{graphicx}
\makeatletter
\usepackage[pdftex]{hyperref}
\usepackage{cite}
\usepackage{etoolbox}

\usepackage{geometry}
\makeatother
\usepackage{wasysym}
\usepackage{tikz}

\makeatletter
\def\@mkboth#1#2{}
\newlength\appendixwidth
\preto\appendix{\addtocontents{toc}{\protect\patchl@section}}
\newcommand{\patchl@section}{%
  \settowidth{\appendixwidth}{\textbf{Appendix }}%
  \addtolength{\appendixwidth}{1.5em}%
  \patchcmd{\l@section}{1.5em}{\appendixwidth}{}{\ddt}%
  \patchcmd{\l@subsection}{2.3em}{\appendixwidth}{}{\ddt}%
}
\makeatother

\usepackage{graphicx,color}
\usepackage[english]{babel}
\usepackage{hyperref}
\usepackage{dsfont}
\renewcommand{\vec}[1]{{\bf #1}}
\newcommand{\eqref}[1]{(\ref{#1})}

\newcommand{\bfr}{ {\bf r}}

\newcommand{\bfu}{ {\bf u}}

\newcommand{\eps}{\varepsilon}

\newcommand{\cF}{\mathcal{F}}

\newcommand{\grad}{\nabla}
\renewcommand{\div}{\nabla\cdot}
\renewcommand{\vec}[1]{\mathbf{#1}}
\newcommand{\Gvec}[1]{\boldsymbol{#1}}

\newcommand\Pe{{\rm Pe}}

\usepackage{upgreek}

\begin{document}
\title[Generalized thermodynamics of Motility-Induced Phase Separation]{Generalized thermodynamics of Motility-Induced Phase Separation: Phase equilibria, Laplace pressure, and change of ensembles.}

\author{Alexandre P. Solon}
\address{Department of Physics, Massachusetts Institute of Technology, Cambridge, Massachusetts 02139, USA}

\author{Joakim Stenhammar}
\address{Division of Physical Chemistry, Lund University, 221 00 Lund, Sweden}

\author{Michael E. Cates}
\address{DAMTP, Centre for Mathematical Sciences, University of Cambridge, Cambridge CB3 0WA, United Kingdom}

\author{Yariv Kafri}
\address{Department of Physics, Technion, Haifa, 32000, Israel}

\author{Julien Tailleur}
\address{Universit\'e Paris Diderot, Sorbonne Paris Cit\'e, MSC, UMR 7057 CNRS, 75205 Paris, France}

\date{\today}

\begin{abstract}
  Motility-induced phase separation (MIPS) leads to cohesive active
  matter in the absence of cohesive forces. We present, extend and
  illustrate a recent generalized thermodynamic formalism which
  accounts for its binodal curve. Using this formalism, we identify
  both a generalized surface tension, that controls finite-size
  corrections to coexisting densities, and generalized forces, that
  can be used to construct new thermodynamic ensembles. Our framework
  is based on a non-equilibrium generalization of the Cahn-Hilliard
  equation and we discuss its application to active particles
  interacting either via quorum-sensing interactions or directly
  through pairwise forces.
\end{abstract}


\maketitle

\tableofcontents
\newpage

One of the most surprising collective behaviors of active particles is
probably the emergence of cohesive active matter in the absence of
cohesive
forces~\cite{Tailleur:2008:PRL,Thompson:2011:JSM,Fily:2012:PRL,Redner:2013:PRL,Bialke:2013:EPL,Stenhammar:2013:PRL,Buttinoni:2013:PRL,Wysocki:2014:EPL,Theurkauff:2012:PRL,Soto:2014:PRE,Wittkowski:2014:NC,Brady:2014:PRL,Speck:2014:PRL,Matas:2014:PRE,Zottl:2014:PRL,Suma:2014:EPL,Solon:2015:PRL,Cates:2015:ARCMP,Redner:2016:PRL,Dijkstra:2016:arxiv,Whitelam:2017:arxiv}. The
underlying linear instability leading to Motility-Induced Phase
Separation (MIPS) is by now well understood~\cite{Cates:2015:ARCMP}:
active particles accumulate where they move more slowly, while
repulsive interactions or steric hindrance slow down active particles
at high density. Active particles thus tend to accumulate where they
are already denser. MIPS has been studied extensively in many
idealized minimal
models~\cite{Tailleur:2008:PRL,Thompson:2011:JSM,Fily:2012:PRL,Redner:2013:PRL,Bialke:2013:EPL,Stenhammar:2013:PRL,Wysocki:2014:EPL}. Most
experimental systems, on the other hand, are too slow or too dilute,
so that only a higher propensity to clustering has been reported in
most cases~\cite{Theurkauff:2012:PRL,Buttinoni:2013:PRL}, with some
notable exceptions~\cite{Liu:2011:SCI,Liu:2017:arxiv}.

While the aforementioned linear instability is well understood,
  and can be used to define a spinodal region, what controls the
  coexisting densities resulting from MIPS has been the topic of a
  long-standing debate. Although the phase coexistence has been mapped
  to an equilibrium
  one~\cite{Tailleur:2008:PRL,Cates:2013:EPL,Speck:2014:PRL,Takatori:2015:PRE},
  this constitutes an ad-hoc approximation that leaves out the
  nonequilibrium contributions specific to MIPS. These have been shown
  to invalidate the equilibrium thermodynamic
  constructions~\cite{Wittkowski:2014:NC,Solon:2015:PRL,Dijkstra:2016:arxiv}
  and thus affect the phase diagram.

  Here we present, complete and extend a recent thermodynamic
  construction for MIPS~\cite{Solon:2016:PRE} starting from a
  non-equilibrium generalization of the Cahn-Hilliard
  equation~\cite{cahn1958free,Bray:2002:AdvPhys} for which we are able
  to compute the coexisting densities analytically. In particular, we
  extend our framework to define a generalized surface tension and
  account for finite-size corrections to coexisting
  densities. Furthermore, our formalism allows us to identify the
  relevant thermodynamic state variables (or generalized forces) which
  can be used to build new thermodynamic ensembles, as we illustrate
  here considering the isobaric ensemble. The macroscopic approach
  described in this article highlights the importance of interfacial
  contributions, which are essential to understand the phase diagram,
  as opposed to the equilibrium case. Moreover, our framework should
  actually be useful beyond MIPS and apply for a larger class of
  non-equilibrium systems exhibiting phase-separation without net mass
  current in the steady state.

The structure of the paper is as follows. First, we consider in
Section~\ref{sec:general} a phenomenological hydrodynamic description
of active systems whose sole hydrodynamic mode is a diffusive
conserved density field. For such systems, we show that the
steady-state configurations---and in particular the phase-separated
profiles---correspond to the extrema of a generalized free energy
functional which we can compute explicitly. As a result, the binodals
are determined at this level from a common tangent construction on a
{generalized} free energy density. Furthermore, we show how our
formalism predicts Laplace-pressure-like corrections to the coexisting
densities for finite systems and define a corresponding generalized
surface tension.

In Section~\ref{sec:QSAPs}, we then consider models in which MIPS
arises from an explicit density-dependence of the propulsion speed
$v(\rho)$~\cite{Tailleur:2008:PRL,Thompson:2011:JSM,Soto:2014:PRE}.
This can be thought of as modeling the way bacteria and other cells
adapt their dynamics to the local density measured through the
concentration of signaling molecules; we refer to such particles as
`quorum-sensing active particles' (QSAPs). We also allow for
anisotropic sensing of the local density field in QSAPs, something
that would be relevant for, \textit{e.g.}, visual cues rather than
chemical ones. We show how, for such models, we can construct a
hydrodynamic description that fits within the framework of
Section~\ref{sec:general}. The latter can then be used to predict
quantitatively the phase diagram of QSAPs and its finite-size
corrections.

In Section~\ref{sec:PFAPs} we then turn to active particles with
constant propulsion forces interacting via isotropic, repulsive
pairwise forces (pairwise force active particles, or
PFAPs)~\cite{Fily:2012:PRL,Redner:2013:PRL,Bialke:2013:EPL,Stenhammar:2013:PRL}. For
these models, the slowdown triggering MIPS is due to
collisions. Contrary to QSAPs, there is no method in the literature
allowing to map the hydrodynamics of PFAPs onto the general framework
of Section~\ref{sec:general}.  Nevertheless, we show that we can still
account for the phase equilibria of PFAPs following the ideas
presented in Section~\ref{sec:general}.

Finally, we show in Section~\ref{sec:Ensemble} how the generalized
thermodynamic variables identified using our formalism play the role
of generalized forces when changing ensembles. In particular, we show
that using an externally imposed mechanical pressure, i.e.,
considering an isobaric ensemble, only leads to a Gibbs phase rule
when mechanical and generalized pressures coincide.

\section{Phase equilibria of a phenomenological hydrodynamic description of MIPS}\label{sec:general}

\subsection{General framework}
\label{sec:generalize-framework}
We consider a continuum description of non-aligning active particles
with isotropic interactions. The vectorial degrees of freedom
corresponding to the particle orientations are then fast degrees of
freedom and do not enter a hydrodynamic description. The sole
hydrodynamic field is thus the conserved density $\rho(\vec r,t)$,
obeying $\dot \rho=-\nabla \cdot {\bf J}$. By symmetry, the current
${\bf J}$ vanishes in homogeneous phases. Its expansion in gradients
of the density involves only odd terms under space reversal. At third
order, we use:
\begin{eqnarray}
  \label{eq:dynamics-general}
\dot\rho &=& \div ( M \grad g[\rho]),\\
  g[\rho]&=&g_0(\rho)+g_1[\rho]\qquad\mbox{where}\qquad g_1=\lambda(\rho) (\grad \rho)^2-\kappa(\rho)\Delta\rho\nonumber.
\end{eqnarray}
Note that for general $\kappa(\rho)$ and $\lambda(\rho)$, $g[\rho]$ cannot be written as the derivative of a free energy. Eq.~\eqref{eq:dynamics-general} is perhaps the simplest generalization of the Cahn-Hilliard equation out of equilibrium and has been argued to be relevant for the phase separation of active particles in the past~\cite{Tailleur:2008:PRL,Stenhammar:2013:PRL,Stenhammar:2014:SM,Wittkowski:2014:NC,Solon:2016:PRE}. For a non-constant $M[\rho]$, it allows for circulating currents with non-zero curls. A generic third order expansion
\begin{equation}\label{eq:currentgeneric}
{\bf J}= \alpha \grad\rho-\kappa \grad \Delta \rho+\lambda \grad
(\grad\rho)^2+[\beta(\grad\rho)^2+\zeta \Delta \rho] \grad \rho
\end{equation}
is formally equivalent to~\eqref{eq:dynamics-general}, at this order
in the gradient expansion, using for instance
$M=1+(\frac\beta\alpha-\frac{\lambda'}{\alpha})(\grad\rho)^2+(\frac\zeta\alpha+\frac{\kappa'}\alpha)\Delta\rho$
and $g_0$ such that $g_0'(\rho)=\alpha(\rho)$, where the prime denotes a derivative with respect to $\rho$. Such choices, however, can lead to a change of sign or a divergence of $M$ so that, in what
follows, we restrict ourselves to dynamics of the
form~\eqref{eq:dynamics-general} with positive definite $M$. Such a restriction does not matter when considering fully-phase separated profiles in the macroscopic limit but was recently proved important when describing curved interfaces~\cite{Tjhung:2018:arxiv} where generic currents of the form~\eqref{eq:currentgeneric} may lead to a richer phenomenology than that of Eq.~\eqref{eq:dynamics-general}.

The spinodal region of a phase-separating system can easily be
predicted from Eq.~\eqref{eq:dynamics-general}. A homogeneous profile
of density $\rho_0$ is indeed linearly unstable whenever
$g_0'(\rho_0)<0$ and the sign of $g_0'(\rho_0)$ hence defines the
spinodal region.

\begin{figure}
  \centering
  \includegraphics[width=0.35\columnwidth]{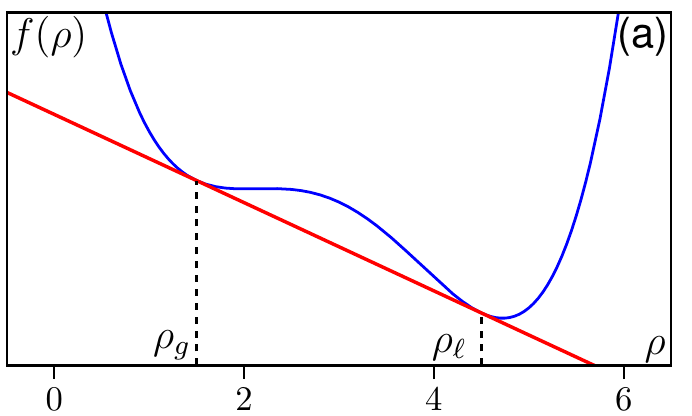}\hspace{0.5cm}
  \includegraphics[width=0.35\columnwidth]{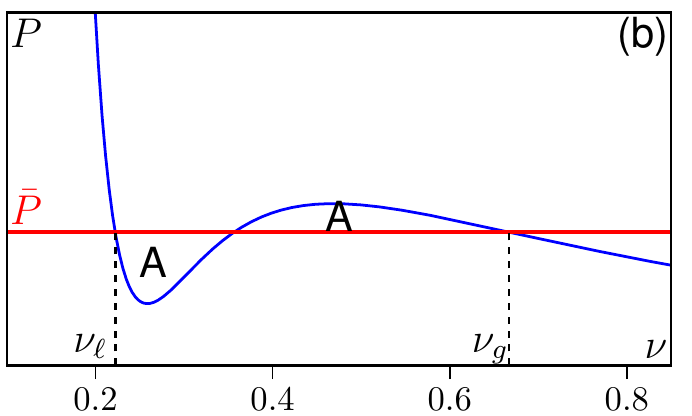}
  \caption{Illustration of the thermodynamic constructions of
    coexisting densities. {\bf (a):}~Common tangent construction on
    the free energy density $f(\rho)$. {\bf (b):}~Maxwell equal-area
    construction on the pressure. In both panels, we used a
    double-well potential for illustrative purpose.}
  \label{fig:equilibrium_constructions}
\end{figure}

\subsection{Warm-up exercise: the equilibrium limit}
Before deriving the binodal curve predicted by
Eq.~(\ref{eq:dynamics-general}) in its most general form, it is illuminating to first review
the corresponding equilibrium limit, i.e., the standard Cahn-Hilliard equation~\cite{cahn1958free,Bray:2002:AdvPhys} which corresponds to $2\lambda+\kappa'=0$~\cite{Solon:2016:PRE}. In this case, the dynamics~\eqref{eq:dynamics-general} corresponds to a steepest descent in a free energy landscape ${\cal F}[\rho]$:
\begin{equation}\label{eq:dyneq}
\dot \rho = \grad \cdot \left[ M \grad \frac{\delta {\cal F}}{\delta \rho} \right]
\qquad\mbox{where}\qquad
{\cal F}[\rho] =\int F[\rho] d{\bf r}= \int \Big[f(\rho)+\frac{\kappa(\rho)}{2}(\grad\rho)^2\Big] d{\bf r}\;.
\end{equation}
$g$ of Eq.~\eqref{eq:dynamics-general} is then the chemical potential,
defined as the functional derivative of ${\cal F}$ with respect to $\rho$:
\begin{equation}\label{eq:FE}
  g=\frac{\delta {\cal F}}{\delta\rho({\bf r})}=g_0(\rho) +g_1[\rho]\;,
\end{equation}
where
\begin{equation}
  g_0(\rho)=f'(\rho)\qquad\mbox{and}\qquad g_1[\rho]=- \frac{\kappa'(\rho)}{2}(\nabla\rho)^2-\kappa(\rho)\Delta\rho\;.
\end{equation}
The free energy functional ${\cal F}$ is extensive so that, in a
macroscopic phase-separated system, the contribution of the interfaces
is sub-dominant. The term $\frac 1 2 \kappa(\rho) (\grad \rho)^2$ in ${\cal F}$ can then be neglected and the phase equilibria can be determined from the bulk free energy density $f(\rho)$: The
coexisting densities $\rho_g$ and $\rho_\ell$ in the gas and liquid
phases are the one minimizing the free energy under the constraint
that the average density $\rho_0$ is fixed. They are obtained through
a common tangent construction on $f(\rho)$ or, equivalently, as the
densities satisfying the equalities of chemical potential
$f'(\rho_g)=f'(\rho_\ell)=\bar\mu$ and pressure
$P(\rho_g)=P(\rho_\ell)=\bar P$, with the pressure $P$ defined as
$P(\rho)=\rho f'(\rho)-f(\rho)$. Alternatively, the coexisting
densities can be constructed using a Maxwell equal area construction
\begin{equation}
  \label{eq:Maxwell-equilibrium}
  \int_{\nu_\ell}^{\nu_g} \left[{P(\nu)-\bar P}\right]d\nu=0
\end{equation}
where $\nu\equiv 1/\rho$ is the volume per particle,
$\nu_{g/\ell}\equiv 1/\rho_{g/\ell}$. The two thermodynamic constructions are illustrated in
Fig.~\ref{fig:equilibrium_constructions}.

Note that, instead of relying on a free energy, the equality of
pressures and chemical potentials between coexisting phases can also
be derived directly from the dynamics~\eqref{eq:dynamics-general}. First the vanishing of the flux $\vec J$ in
Eq.~\eqref{eq:dynamics-general} immediately imposes a uniform chemical
potential $g$, which is thus equal between coexisting phases:
$g_0(\rho_\ell)=g_0(\rho_g)$.  To derive the equality of pressures, we
rewrite Eq.~\eqref{eq:dynamics-general} as
\begin{equation}\label{eq:sigmaeq}
	\dot \rho = - \nabla \cdot \left[ \frac{M}{\rho} \nabla \cdot {\Gvec \sigma} \right],
\end{equation}
where ${\Gvec \sigma}$ is the stress tensor, whose expression in Cartesian
coordinates is
\begin{equation}\label{eq:sygmaeqexpr}
\sigma_{\alpha\beta} = - \delta_{\alpha\beta}\left[ P(\rho) -\frac{\rho \kappa'+\kappa}2 (\nabla \rho)^2-\rho \kappa \Delta \rho \right]-\kappa (\partial_\alpha \rho)(\partial_\beta \rho)\;.
\end{equation}
Note that, similar to $g$, ${\Gvec \sigma}$ is related to the free-energy
functional through~\cite{Kruger2017}:
\begin{equation}
  {\sigma}_{\alpha\beta}=\delta_{\alpha\beta} \left[F-\rho\frac{\delta {\cal F}}{\delta \rho} \right]-\frac{\partial F}{\partial (\partial_\beta\rho)}\partial_\alpha\rho.
\end{equation}
In fully phase-separated, flux-free steady states, one can get the equality of pressure between coexisting homogeneous phases from Eqs.~\eqref{eq:sigmaeq}-\eqref{eq:sygmaeqexpr}.
For finite systems, Eq.~\eqref{eq:sigmaeq} can also be  used to derive the finite-size corrections to the binodals due to Laplace pressure~\cite{Bray:2002:AdvPhys}.

\subsection{Generalized thermodynamic variables}
\label{sec:generalized-variables}
For generic functions $\lambda(\rho)$ and $\kappa(\rho)$, which do not
satisfy $2\lambda(\rho)+\kappa'(\rho)=0$, the free energy structure
breaks down because the gradient terms in $g$ cannot be written as a functional derivative:
\begin{equation}
	g_1[\rho]=\lambda(\rho) (\grad \rho)^2-\kappa(\rho)\Delta\rho \neq \frac{\delta {\cal F}}{\delta \rho}.
\end{equation}
A common tangent construction on a free energy density defined through
$f'(\rho)=g_0(\rho)$ then does not lead to the correct coexisting
densities~\cite{Wittkowski:2014:NC,Solon:2016:PRE}. However, as we
show below, $g$ can be written as the functional derivative of a
generalized free energy ${\cal G}$ with respect to a non-trivial new
variable $R$, which depends on the functional forms of $\kappa$ and
$\lambda$.  Although the dynamics~\eqref{eq:dynamics-general} are
\textit{a priori} out of equilibrium, its steady states correspond to extrema of this
generalized free energy and, as we show below, we recover the full structure of the
equilibrium case described above. We now derive this mapping and show
in Section~\ref{sec:binodals} how it can be used to compute the
binodals of Eq.~(\ref{eq:dynamics-general}) exactly. Finally, we turn
to their finite-size corrections in Section~\ref{sec:FFS-GT}.

To proceed, we consider the one-to-one mapping $R(\rho)$ defined by
\begin{equation}\label{eqR}
  \kappa R'' = -(2\lambda + \kappa') R'\;,
\end{equation}
where the derivatives are taken with respect to $\rho$. Direct
inspection shows that $g$ can now be written as a functional
derivative with respect to $R$~\cite{Solon:2016:PRE}:
\begin{equation}
g=\frac{\delta {\cal G}}{\delta R}
\end{equation}
with
\begin{equation}\label{eq:generalizedF}
  {\cal G}=\int d {\vec r}\, G[R] \equiv \int d {\vec r} \left[\phi(R) +\frac{\kappa}{2 R'}
    (\grad R)^2\right]
\end{equation}
where we have defined a generalized free energy density $\phi(R)$ such
that
\begin{equation}
  \label{eq:phi-definition}
  \frac{d\phi}{dR}=g_0\;\qquad\mbox{or alternatively}\qquad \phi = \int^\rho g_0(\hat \rho) R'(\hat \rho) d \hat \rho\; .
\end{equation}
The dynamics of $\rho$ is now written as the derivative of a
generalized free energy functional:
\begin{equation}\label{eq:generalizeddyn}
  \dot \rho= \grad \cdot \left[ M[\rho] \grad \frac{\delta {\cal G}}{\delta R} \right]\;.
\end{equation}
Note, however, that the structure of~\eqref{eq:generalizeddyn}
differs from the equilibrium case~(\ref{eq:FE}) since the
functional derivative is taken with respect to 
$R$ instead of $\rho$. Nevertheless, the steady-state solutions of~\eqref{eq:generalizeddyn} correspond to extrema of ${\cal G}$ with respect to $R$~\footnote{The dynamics of $R$ itself can be easily deduced as $\dot R = R' \;\grad \cdot [M \grad \frac{\delta {\cal G}}{\delta R}]$. Note that, in particular, $R$ is not a conserved quantity. }.

Comparing Eq.~\eqref{eq:generalizeddyn}
to the equilibrium case~\eqref{eq:dyneq}, we note that the
former can be seen as driven by gradients
of a generalized chemical potential $g=\frac{\delta {\cal G}}{\delta R}$. Similarly to the equilibrium
case, we now show that the dynamics~(\ref{eq:generalizeddyn}) can 
also be written so as to appear driven by the divergence of a
generalized stress tensor. Specifically, the current $\vec J$ can be
rewritten as
\begin{equation}
  \label{eq:current-stress}
  \vec J=-M\grad g=\frac{M}{R}\div \Gvec\sigma
\end{equation}
with a tensor $\Gvec\sigma$ reading in Cartesian coordinates
\begin{eqnarray}
  \label{eq:stress-tensor}
  \sigma_{\alpha\beta} &=& -\left[h_0+Rg_1-\frac{\kappa R'}{2}(\grad\rho)^2\right]\delta_{\alpha\beta}-\kappa R' (\partial_\alpha\rho)(\partial_\beta\rho)\;,
\end{eqnarray}
where we have defined~\footnote{Alternatively, $h_0$ can be obtained
  through $ h_0 = \int^\rho R(\hat\rho) g_0'(\hat\rho) d\hat\rho$, or,
  introducing $\upupsilon=1/R$, through
  $h_0=-\frac{d (\phi \upupsilon)}{d \upupsilon}$.}
\begin{equation}\label{eq:h0def}
   h_0 = R\frac{d\phi}{dR}-\phi\;.
\end{equation}
Once again, the generalized stress tensor can be deduced from the
generalized free energy through
\begin{equation}
  \sigma_{\alpha\beta}=\delta_{\alpha\beta} \left[G-R\frac{\delta {\cal G}}{\delta R} \right]-\frac{\partial G}{\partial (\partial_\beta R)}\partial_\alpha R\;.
\end{equation}
In the following, we identify the diagonal coefficients of
${\Gvec \sigma}$, the normal stresses, with generalized (potentially
anisotropic) pressures. Again, we split $h=-\sigma_{xx}$ into a local
function and an interfacial contribution:
\begin{equation}\label{eq:h0h1}
h=h_0(\rho)+h_1[\rho]\qquad \mbox{where}\qquad  h_1= Rg_1 - \frac{\kappa R'}{2}(\grad\rho)^2+\kappa R' (\partial_x \rho)^2\;.
\end{equation}
We emphasize here that $\Gvec\sigma$ and $h$ need not have any
connection to mechanics and momentum transfer.

Finally, we stress that the equilibrium case is easily recovered using
$2\lambda + \kappa'=0$: Eq.~(\ref{eqR}) then implies that $R=\rho$ (up
to multiplicative and additive constants that play no role in phase
equilibria and can thus be discarded). All our generalized quantities
then reduce to their equilibrium counterparts. 

Before we turn to the derivation of the binodal curve, we first note that
\begin{equation}
\frac{d^2\phi}{d R^2}=\frac{g_0'(\rho)}{R'(\rho)}\;.
\end{equation}
The spinodal region, defined as $g_0'(\rho)<0$, thus corresponds to
the region in which the generalized free energy density is concave,
$\frac{d^2\phi}{d R^2}<0$, provided $R'$ is chosen
positive. Furthermore, from Eq.~\eqref{eq:h0def} one finds that
\begin{equation}
h_0'(\rho)=R g_0'(\rho)
\end{equation}
so that the spinodal region can equivalently be defined from
$h_0'(\rho)<0$. Finally, we note that, contrary to the generalized
free energy density $\phi$ which depends on $\lambda$ and $\kappa$
through $R$, the spinodal region is unaffected by the gradient terms
in $g$.

We now show how the above results directly yield the binodal curve of
our generalized Cahn-Hilliard equation by considering fully
phase-separated systems. We then discuss in Section~\ref{sec:FFS-GT}
the corrections to the binodal curve for finite-size systems.

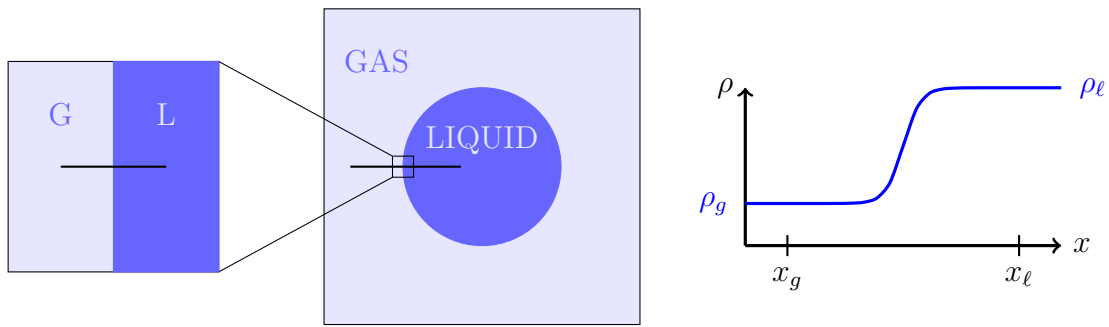
\begin{figure}
  \centering
  \begin{tikzpicture}[scale=1.4]
    \draw[black,fill=blue!10!white] (0,0) rectangle (3,3);
    \draw[blue!60!white] (0.5,2.5)  node {GAS};
    \filldraw[blue!60!white] (1.5,1.5) circle (.75);
   
    \draw[blue!10!white] (1.5,1.75)  node {LIQUID};
    \draw[thick] (.25,1.5) -- (1.3,1.5);
	\draw[black] (.75-.1,1.5-.1) rectangle (.75+.1,1.5+.1);
    \draw[black] (.75-.1,1.5-.1) -- (-1,.5);
    \draw[black] (.75-.1,1.5+.1) -- (-1,2.5);
	\draw[black,fill=blue!10!white] (-3,.5) rectangle (-1,2.5);
    \filldraw[blue!60!white] (-2,.5) rectangle (-1,2.5);
    \draw[blue!60!white] (-2.5,2)  node {G};
    \draw[blue!10!white] (-1.5,2)  node {L};
    \draw[thick] (-2.5,1.5) -- (-1.5,1.5);	
    
    \begin{scope}[xshift=4cm,yshift=.75cm]
      \draw[->,line width=1.2pt] (0,0) -- (3,0) node[right] {$x$};
      \draw[->,line width=1.2pt] (0,0) -- (0,1.5) node[left] {$\rho$};
      \draw[scale=1,domain=0:3,smooth,variable=\x,blue,line width=1.2pt] plot ({\x},{0.4+0.55*(tanh((\x-1.5)*6.)+1))});
      \draw[blue] (-0.3,0.4) node {$\rho_g$};
      \draw[blue] (3.3,1.5) node {$\rho_\ell$};
      \draw[line width=0.8pt] (0.4,0.1) -- (0.4,-0.1) node [below] {$x_g$};
      \draw[line width=0.8pt] (2.6,0.1) -- (2.6,-0.1) node [below] {$x_\ell$};
    \end{scope}
  \end{tikzpicture}
  \caption{Schematic representation of the mean density field of a
    fully phase-separated system in 2d. We consider the density
    profile connecting gas and liquid phases along a horizontal cut so
    that the interface is oriented along ${\bf \hat y}$ (center). 
    In the macroscopic limit, the interface is locally
    flat in the transverse direction $\bf \hat y$ (left) and the
    problem simplifies into an effectively 1d domain wall computation
    for the density profile (right).} \label{fig:scheme}
\end{figure}

\subsection{Phase coexistence in the large system size limit}
\label{sec:binodals}
A macroscopic droplet of, say, the dense phase has an infinite radius
of curvature in the large system size limit, so that curvature effects
are negligible. As in equilibrium, computing the coexisting densities reduces to studying a one-dimensional domain-wall profile perpendicular to the interface~\cite{Bray:2002:AdvPhys}, whatever the original number of
spatial dimensions. To do so, we
consider a flat interface, orthogonal to $\mathbf{\hat{x}}$, between
coexisting gas and liquid phases at densities $\rho_g$ and $\rho_\ell$
(see Fig.~\ref{fig:scheme}).

For such a profile, any derivative with respect to a direction normal to $\hat {\vec x}$ vanishes so that Eq.~\eqref{eq:current-stress} directly implies that $g$ and $\sigma_{xx}$ are constant. For coexisting homogeneous phases, this  leads directly to
\begin{equation}\label{eq:binodal_cond}
g_0(R_\ell)=g_0(R_g)\qquad\mbox{and}\qquad h_0(R_\ell)=h_0(R_g)
\end{equation}
where $R_{\ell,g}\equiv R(\rho_{\ell,g})$. These two constraints thus fully determine the coexisting densities and are equivalent to a common tangent construction on $\phi(R)$ since $g_0=\frac{d \phi}{d R}$ and $h_0=R \frac{d \phi}{d R} -\phi$.

In stark contrast to equilibrium liquid-gas phase separation, the
interfacial terms $g_1$ or $h_1$ affect the coexisting densities
through the definition of $R$, Eq.~(\ref{eqR}), which depends on
$\lambda(\rho)$ and $\kappa(\rho)$. Note that the sole knowledge of
the dynamics in Eq.~\eqref{eq:dynamics-general} allows us to
determine the coexisting densities using the constructions above,
without the need to solve for the full density profile at the
interface. The 
common-tangent construction on $\phi$ leads to coexisting densities
which are independent of the mean density $\rho_0$. The lever rule for
determining the phase volumes $V_\ell$ and $V_g$ therefore still
applies: $\rho_\ell V_\ell+\rho_g V_g = \rho_0 V$. Note that this
lever rule applies to $\rho$ and not to $R$ since the latter is not a
conserved quantity.

{\it The Maxwell construction.} As in equilibrium, the common tangent
construction on $\phi$ is equivalent to a Maxwell construction on
$h_0$. We now derive the latter because it will be useful when
considering PFAPs, and also since it provides a simpler numerical route
to computing the binodal curve from the expression of $h$.

As we shall do for PFAPs, we start from a current given by
Eq.~\eqref{eq:current-stress} so that the flux free condition in a
situation as depicted in Fig.~\ref{fig:scheme} implies that the
generalized pressure is constant, recalling that the curvature of the interface is negligible:
\begin{equation}
  h=h_0 + h_1 = \bar h
\end{equation}
Then, we introduce the
generalized volume per particle
\begin{equation}
\upupsilon=\frac 1 R
\end{equation}
and compute the integral
\begin{equation}\label{eq:MCP}
\int_{\upupsilon_\ell}^{\upupsilon_g} (h_0-\bar h) d \upupsilon= -\int_{x_g}^{x_\ell}  (h_0-\bar h) \partial_x\upupsilon dx = \int_{x_g}^{x_\ell} h_1  \partial_x\upupsilon dx
\end{equation}
where the spatial integral is computed along the direction normal to
the interface. After some algebra,  $h_1$ can be rewritten as
\begin{equation}
  h_1=\frac 1 {R'} \left[ \left( \kappa-\frac{R\kappa'}{R'}\right) \frac{(\partial_x
    R)^2}{2}-R\kappa \partial_{xx} R \right]+\frac{R \kappa R''}{2
    (R')^3}(\partial_x R)^2.
\end{equation}
This allows us, after some
algebra, to show that $h_1\partial_x \upupsilon$ is a total derivative
\begin{equation}
h_1  \partial_x \upupsilon = \partial_x \left[\frac{\kappa (\partial_xR)^2}{2R R'}\right]\;.
\end{equation}
In turn, this leads to a generalized Maxwell construction on $h_0$:
\begin{equation}\label{eq:MCP}
\int_{\upupsilon_\ell}^{\upupsilon_g} (h_0-\bar h) d \upupsilon=0\;.
\end{equation}

\subsection{Finite size effects}
\label{sec:FFS-GT}
Let us now consider what happens if one takes into account the finite
curvature of the phase-separated domains. Again, thanks to our
mapping, the derivation below resembles closely the one done in equilibrium for the  Cahn-Hilliard equation~\cite{Bray:2002:AdvPhys}. We consider a radial cut along the
interface of a circular domain in 2D, as in Fig~\ref{fig:scheme}. By
symmetry, the current $\vec J$ vanishes in steady state.
Eq.~(\ref{eq:current-stress}) then immediately gives $\grad g=0$ so
that one still has an equality of generalized chemical potentials
between the two phases: $g_0(\rho_g)=g_0(\rho_\ell)$. On the other
hand, $\div \Gvec\sigma=\vec 0$ does not lead to a uniform
$\sigma_{xx}$ in this circular geometry, which highlights the
different behaviors of the generalized chemical potential and the
generalized stress tensor for finite systems.

To proceed, we integrate the radial component of $\div \Gvec\sigma$ along
the path depicted in Fig.~\ref{fig:scheme}. To highlight the spherical geometry, we parametrize this path as $r \vec{\hat r}$. Using the expression for
the divergence of a tensor in spherical coordinates (polar in 2D) leads
to
\begin{equation}
  \int_{r_\ell}^{r_g} (\div \Gvec\sigma) \cdot \vec {\hat r} dr = 0 =    \int_{r_\ell}^{r_g} \left[\partial_r \sigma_{rr} + \frac{1}{r} (\sigma_{rr}-\sigma_{\theta\theta})\right]dr
\end{equation}
Using the expression~\eqref{eq:stress-tensor} of $\Gvec \sigma$ in this geometry then leads to
\begin{equation} \label{eq:halas}
\sigma_{rr}(r_\ell)-\sigma_{rr}(r_g)=h_0(\rho_g)-h_0(\rho_\ell)=  -  \int_{r_\ell}^{r_g}  \left[\frac{R'}{r} \kappa (\partial_r \rho)^2 dr\right]\;,
\end{equation}
where we have used that the isotropic terms in $\Gvec \sigma$ cancel
and derivatives with respect to $\theta$ vanish by symmetry.

When the width of the interface is small compared to the droplet
radius $r_d$, expanding $r$ around $r_d$ and using that
$(\partial_r\rho)^2$ vanishes outside the interface leads to
\begin{equation}
  \Delta h_0\equiv h_0(\rho_\ell)-h_0(\rho_g)\simeq \frac\gamma{r_d}\;,\label{eq:pressurejump}
\end{equation}
where we have introduced a generalized surface tension $\gamma$:
\begin{equation}\label{eq:surftensgen}
\gamma=\int^{r_g}_{r_\ell}   R' \kappa (\partial_r \rho)^2 dr\;.
\end{equation}
Note that, as for $h_0$ and $\sigma$, $\gamma$ need not have any
mechanical interpretation for generic phase-separating active matter
systems. To leading order in $1/r_d$, $\gamma$ can be computed across
a flat interface (using a slab geometry as in
Fig.~\ref{fig:scheme}). For an interface perpendicular to the
$x$-axis, it then reads
\begin{equation}
  \label{eq:delta-straight}
  \gamma=\int_{x_g}^{x_\ell} (\sigma_{yy}-\sigma_{xx}) dx=\int_{x_g}^{x_\ell} R' \kappa (\partial_x \rho)^2 dx.
\end{equation}

Finally, let us comment on the sign of $\gamma$ which has recently
attracted interest since it has been measured negative for
PFAPs~\cite{Bialke:2015:PRL} (see Section~\ref{sec:PFAP-FS} for a
discussion of that case). Here, since $\kappa$ need to be positive for
stability reasons, we see from Eq.~(\ref{eq:delta-straight}) that
$\gamma$ has the sign of $R'$. Starting from the dynamics in
Eq.~(\ref{eq:dynamics-general}), the sign of $R'$ is arbitrary,
corresponding to an integration constant when solving
Eq.~(\ref{eqR}). The generalized surface tension $\gamma$ can then be
either positive or negative, although with different expressions for
the generalized pressure $h_0(\rho)$. On the other hand, starting from
an expression for the stress tensor in Eq.~(\ref{eq:stress-tensor}), as
will be the case for PFAPs in Section~\ref{sec:PFAPs}, $R'$ is fixed by
the expression for $\Gvec\sigma$ and can take either sign. Our
framework thus supports both positive and negative $\gamma$.


\subsection{Illustration of our general framework for a scalar active matter model}

\label{sec:PDEexample}

\begin{figure}
  \centering
  \includegraphics[width=0.5\columnwidth]{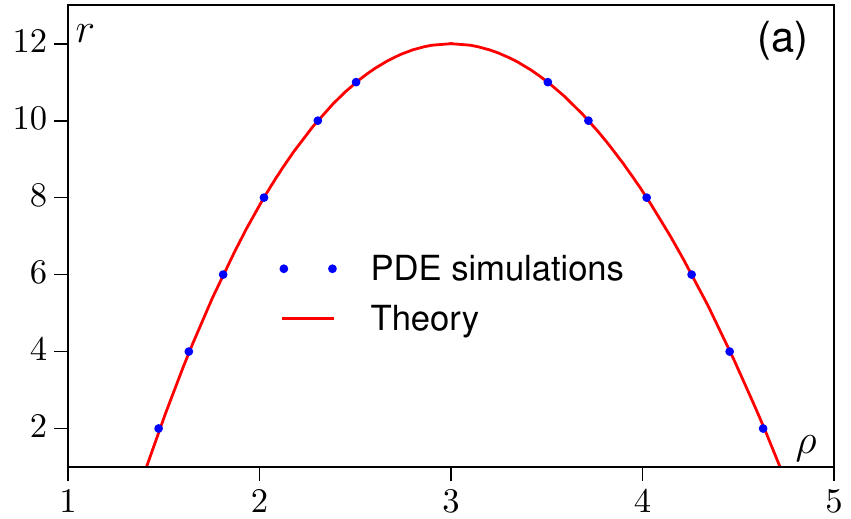}
  \raisebox{1cm}{\includegraphics[width=0.24\columnwidth]{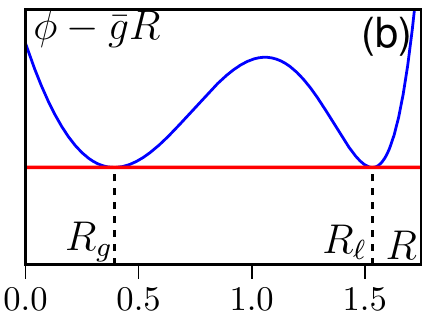}\includegraphics[width=0.24\columnwidth]{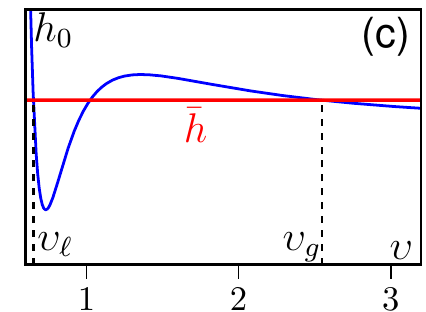}}
  \caption{{\bf (a):} Binodals measured in 2d simulations of {Eq. \eqref{eq:dynamics-general}} with the coefficients of
    Eq.~\eqref{eq:CHex} compared to the theoretical predictions of
    Eq.~\eqref{eq:binodal_cond}. {\bf (b):} Common tangent
    construction on $\phi(R)$ for $r=2$. {\bf (c):} Maxwell construction on $h_0(\upupsilon)$,
    Eq.~(\ref{eq:MCP}) for $r=2$. Eq.~\eqref{eq:dynamics-general} was integrated
    with a precision of $dx=1$ in space and a time step
    $dt=2.5\times 10^{-4}$ for a system size of $100 \times 20$.}
  \label{fig:PDE_phase-diagram}
\end{figure}

In this section we show on a particular example that our generalized
thermodynamic construction
predicts exactly the phase equilibria of our nonequilibrium
Cahn-Hilliard equation~\eqref{eq:dynamics-general} through Eq.~\eqref{eq:binodal_cond}. To this end, we
numerically integrate this equation in 2d for the particular (and
rather arbitrary) choice:
\begin{equation}\label{eq:CHex}
  g_0(\rho)=r(\rho-\rho_0)-12(\rho-\rho_0)^2+4(\rho-\rho_0)^3 \,; \quad M=1\,;  \quad \kappa(\rho)=\rho \,; \quad \lambda=0.
\end{equation}

To check the theory, we first numerically solve
Eqs.~\eqref{eq:dynamics-general} and~\eqref{eq:CHex} using
a semi-spectral integration scheme (linear terms are computed in
Fourier space, non-linear terms in real space) with Euler time
stepping. For each value of $r$, we start from a phase-separated state
with two arbitrarily chosen densities ($\rho_g=1$ and $\rho_\ell=5$)
and measure the coexisting densities once the system has relaxed to
its steady state.

To compare with the theoretical predictions for the binodals, we first
determine the function $R(\rho)$ using Eq.~\eqref{eqR}, which (up to two unimportant integration constants)
gives $R(\rho)=\log \rho$. We then use either the
common tangent construction on $\phi(R)$ or the Maxwell construction
on $h_0(\upupsilon)$, shown in Fig.~\ref{fig:PDE_phase-diagram}(b,c),
as described in the previous section. The comparison with the coexistence densities
measured in the simulations of Eq. \eqref{eq:dynamics-general} is shown in Fig.~\ref{fig:PDE_phase-diagram}(a): the
difference between theory and simulations is found to be smaller than
$0.5$\% for every point, thus confirming that the dynamics does indeed yield 
the stationary state analyzed in Section~\ref{sec:general}.

\begin{figure}
\begin{center}
\includegraphics[width=0.48\columnwidth]{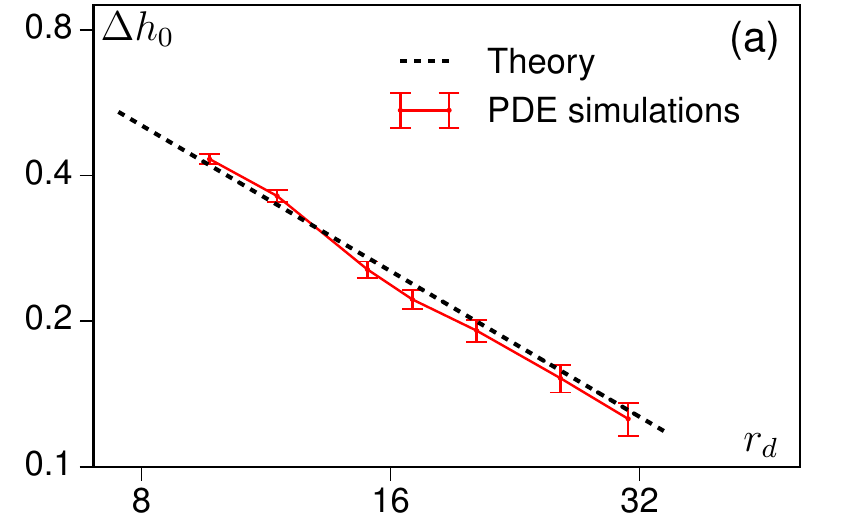}\hspace{0.3cm}
\includegraphics[width=0.48\columnwidth]{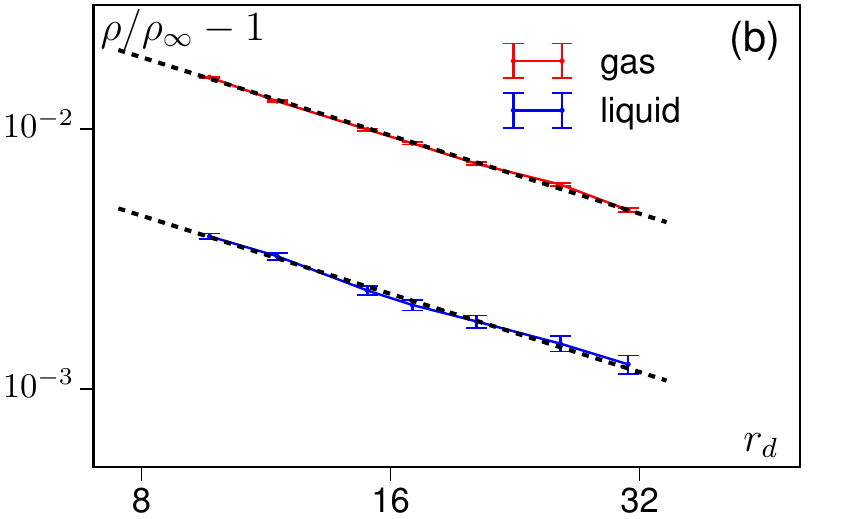}
\end{center}
\caption{Finite size effects measured in 2d
  simulations of Eqs. \eqref{eq:dynamics-general} and \eqref{eq:CHex} with $r=2$. {\bf (a):} {Difference in the
  generalized pressure $h_0$ between the two phases.} The dashed line is the predicted leading order
  behavior $\gamma/r_d$ with the effective surface tension
  $\gamma=4.06$ measured independently from a straight interface using
  Eq.~(\ref{eq:surftensgen}). {\bf (b):} Corrections to the
  coexisting densities. The dashed lines are fits to $c_{g,\ell}/r_d$ with $c_{g,\ell}$ a phase dependent constant. 
  The system size is $80\times 80$, and the other parameters as in Fig. \ref{fig:PDE_phase-diagram}.}\label{fig:FFS-PDE}
\end{figure}

These numerical results are obtained in systems where a straight band
of liquid coexists with a dilute gas phase so that finite-size curvature effects are negligible. On the contrary, when finite-size liquid droplets coexist with a gaseous background, the coexisting densities differ from those predicted by Eq.~\eqref{eq:binodal_cond} due to the finite-size corrections discussed in Section~\ref{sec:FFS-GT}. In this case, a jump of the generalized pressure through the interface is indeed measured numerically, and found to be given quantitatively by the generalized surface tension~\eqref{eq:surftensgen} (See Fig.~\ref{fig:FFS-PDE}a). Similarly,
there are density shifts in each of the phases which scale as $1/r_d$ as shown in Fig.~\ref{fig:FFS-PDE}b.

\section{QSAPs.}\label{sec:QSAPs}
We now turn to a microscopic model for QSAPs, for which we derive a hydrodynamic description and compare the predictions of our formalism
with direct numerical simulations of the microscopic model. We consider particles labeled by $i=1 \dots N$, 
moving at speed $v$ along body-fixed directions $\vec u_i$ which undergo both continuous
rotational diffusion with diffusivity $D_r$ and complete
randomization with tumbling rate $\alpha$. The equations of motion are given by the Langevin dynamics
\begin{eqnarray}
  \label{eq:def-QSAPs}
  \dot{\bfr}_i &=v\left[\tilde\rho (\bfr_i+\eps \bfu_i)\right]\bfu_i +\sqrt{2D_t}\Gvec\eta_i \\
  \dot \theta_i &=\sqrt{2D_r}\xi_i + \sum_j \delta(t-t_j) \delta\theta_j \nonumber
\end{eqnarray}
where $\Gvec\eta$ and $\xi$ are delta-correlated Gaussian white noises of appropriate dimensionality. In addition to continuous angular diffusion, we have included in~\eqref{eq:def-QSAPs} a non-Gaussian noise accounting for tumbling events: the $t_i$ are Poisson distributed with a rate $\alpha$ and the $\delta\theta_j$'s are drawn from a uniform distribution between $0$ and $2\pi$. Each particle adapts its
speed, $v[\tilde\rho(\vec r_i+\eps \vec u_i)]$, to a local measurement
of the density:
\begin{equation}  \label{eq:rho-kernel}
  \tilde \rho(\vec r+\eps \vec u_i)=\int d\vec r' K(\vec r+\eps \vec u_i-\vec r')\hat\rho(\vec r') d\vec r'
\end{equation}
with $K(\vec r)$ an isotropic coarse-graining kernel, and
$\hat\rho(\vec r)=\sum_i \delta(\vec r-\vec r_i)$ the microscopic
particle density. Note that the local density is measured with an
offset $\eps \vec u_i$ which allows for \emph{anisotropic} quorum
sensing. This effect, which does not create alignment interactions,
captures a slowdown of particles that would arise, for instance, due
to a large density of particles \textit{in front} of them. This can
thus model, say, a visual quorum-sensing or steric hindrance. In a
different context, anisotropic sensing has been shown to lead to a
rich phenomenology for aligning active
particles~\cite{chen_fore-aft_2017}.

\subsection{Hydrodynamic description of QSAPs.}
Deriving hydrodynamic equations from microscopics is generally
difficult, even in equilibrium~\cite{Kipnis:2013:book}. For QSAPs we
can follow the path
of~\cite{Tailleur:2008:PRL,Cates:2013:EPL,Solon:2015:EPJST}, taking a
mean-field approximation of their \textit{fluctuating} hydrodynamics.
We first assume a smooth density field and a short-range anisotropy so
that the velocity can be expanded as
\begin{equation}
  \label{eq:expansion-v}
  v(\tilde\rho) \simeq v(\rho)+\eps v'(\rho)\grad\rho\cdot\vec{u}_i+\ell^2v'(\rho)\Delta\rho+{\cal O}(\eps^2,\grad^3)
\end{equation}
where $\rho$ is evaluated at $\bfr_i$ and
$\ell^2=\int r^2 K(\vec r)d\vec
r$.
Following~\cite{Cates:2013:EPL,Solon:2015:EPJST}, the fluctuating
hydrodynamics of QSAPs, derived
in~\ref{app:drif-diffusion}, is then given by:
\begin{equation}
  \dot\rho=\grad \cdot (M \grad g + \sqrt{2 M \rho} {\bf\Lambda})
\end{equation}
with ${\bf\Lambda}$ a unit Gaussian white noise vector and 
\begin{eqnarray}\label{eq:paramMF}
  g_0(\rho)=\log(\rho v)+\frac{\eps}{\tau v};\quad
  M=\rho\frac{\tau v(\tilde\rho)^2}{d};\quad
  \kappa(\rho)=-\ell^2 \frac{v'}{v}\left(1-\frac{\eps}{\tau v}\right);\quad \lambda(\rho)=0\;,
\end{eqnarray}
where $d$ is the number of spatial dimensions. Here, $\tau \equiv [(d-1)D_r+\alpha]^{-1}$ is the orientational persistence
time. The mean-field hydrodynamic equation of QSAPs is then
Eq.~\eqref{eq:dynamics-general} {with the coefficients in}
Eq.~\eqref{eq:paramMF}. As mentioned earlier, the spinodal region is defined from the criterion $g_0'(\rho)<0$, which leads here to a modification of the standard linear instability criterion for QSAPs~\cite{Tailleur:2008:PRL}:
\begin{equation}
\frac{v'(\rho)}{v(\rho)}\left(1-\frac{\eps}{\tau v}\right)<-\frac 1 {\rho}\;.
\end{equation}

To construct the phase diagram for a given choice of $v(\rho)$, using
the generalized thermodynamic procedure, we first solve for $R(\rho)$
using Eq.~(\ref{eqR}) and from it obtain both $\phi(R)$ and
$h_0(R)$. The binodals then follow via a common-tangent construction
on $\phi(R)$ or, equivalently, by setting equal values of $h_0$ and
$g_0$ in {coexisting} phases. Note that since $2\lambda+\kappa'\neq 0$
one has $R \neq \rho$. The phase diagram thus cannot be found by
globally {minimizing} a free energy density $f(\rho)$ defined from
$f'(\rho) = g_0(\rho)$ as discussed
before~\cite{Tailleur:2008:PRL,Cates:2013:EPL}. Indeed, such a
procedure correctly captures the equality of $g_0$ in both phases but
predicts a common tangent construction on $f$ which is violated. We
now turn to describe the numerical simulations of microscopic models
of QSAPs.

\subsection{Comparison between theory and numerics.}

In what follows we study models where the density $\tilde\rho$ is computed according to Eq.~(\ref{eq:rho-kernel}) with the bell-shaped kernel 
\begin{equation}
  \label{eq:kernel}
  K(r)=\frac{\Theta(r_0-r)}{Z} \exp \left( -\frac{r_0^2}{r_0^2-r^2} \right) \;.
\end{equation}
Here $\Theta$ is the Heaviside function, $Z$ a normalization constant and we used  an interaction radius of $r_0=1$. In addition we take the velocity to be
\begin{equation}
  \label{eq:vrho-tanh}
  v(\rho)=v_0+\frac{v_1-v_0}{2}\left[1+\tanh\left(2\frac{\rho}{\rho_m}-2\right)\right] \;.
\end{equation}
This interpolates smoothly between a high velocity $v_0$ at low density ($\rho\ll\rho_m$) and a low velocity $v_1$ at high density
($\rho\gg \rho_m$). In addition to the 2d continuous space model described above, we also conducted simulations of QSAPs in 1d on  lattice~\cite{Thompson:2011:JSM}. In this case, we consider run-and-tumble particles (RTPs): particle $i$ has a direction of motion $u_i=\pm 1$ which is flipped at rate $\alpha/2$. It then jumps on the lattice site in direction $u_i$ with rate $v[\tilde \rho(x_i+\eps u_i)]$. 

\begin{figure}
  \centering
  \includegraphics[width=0.45\columnwidth]{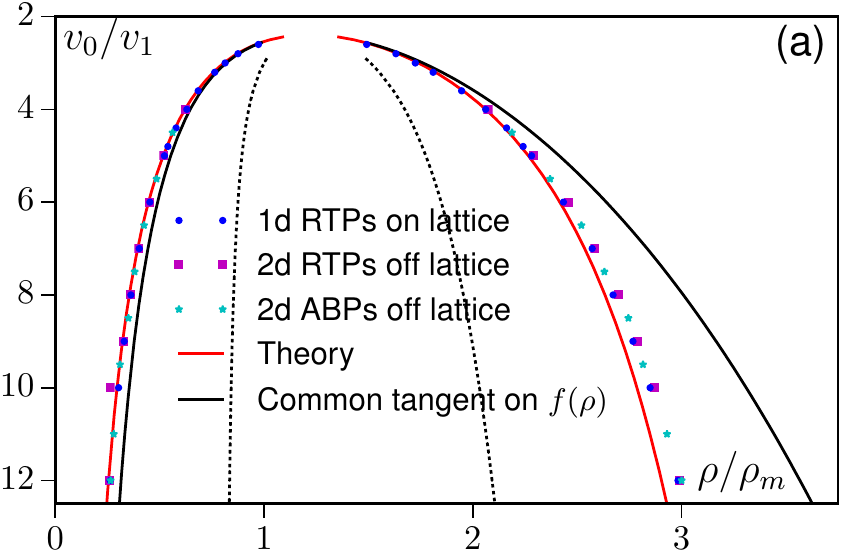}\hspace{0.2cm}
  \includegraphics[width=0.45\columnwidth]{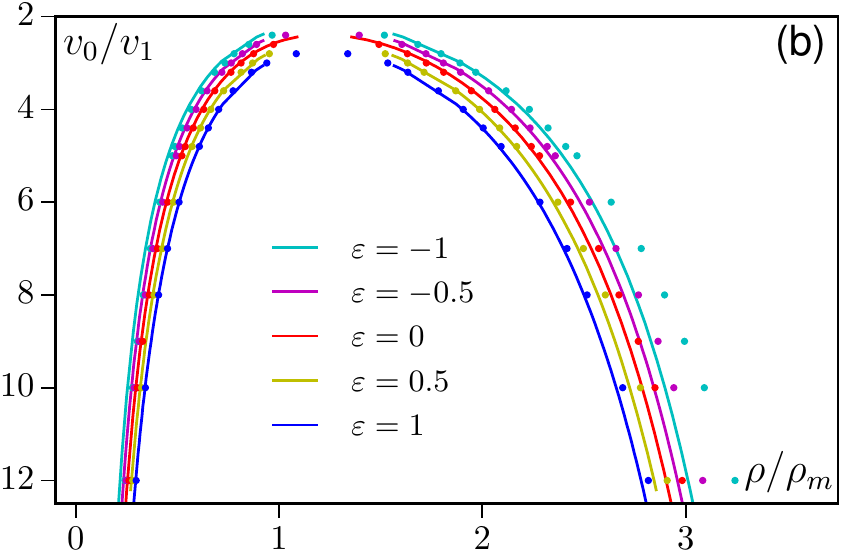}
  \caption{Phase diagrams of QSAPs. {\bf (a)} Symmetric sensing
    ($\eps=0$). The solid lines correspond to common tangent
    constructions on $\phi(R)$ (red) or $f(\rho)$ (black). Dashed
    lines correspond to the spinodals $d^2\phi/dR^2 = d^2f/d\rho^2=0$.
    Data points are from simulations of RTPs ($\alpha=1,\,D_r=0$) and
    ABPs ($\alpha=0,\,D_r=1$), either in 1d on lattice (system size
    $L=2000$ lattice sites) or in a 2d continuous space (system size
    $50\times 50$).  {\bf (b)} Asymmetric sensing ($\eps \neq 0$) for
    1d RTPs on lattice: Solid lines show the predicted binodals
    computed by common tangent constructions on $\phi(R)$, and symbols
    denote simulation results obtained with systems of size $L=2000$
    lattice sites.  For all plots, we used $\rho_m=200$, $v_1=5$,
    $\tau=1$.}
  \label{fig:vrho}
\end{figure}

Fig.~\ref{fig:vrho}a shows the phase diagrams predicted by our
generalized thermodynamics and those measured in QSAP simulations for a
symmetric sensing of the density ($\eps=0$). Overall, the agreement
between predicted and measured binodals is excellent, in contrast to
the common tangent construction on $f(\rho)$. It is remarkable that,
for QSAPs, we can quantitatively predict the phase diagram of a
microscopic model without any fitting parameters, something rare even
for equilibrium models.

Fig.~\ref{fig:vrho}b shows the binodals measured in 1d
simulations on lattice with $\eps\neq 0$ together with the corresponding
theoretical predictions. The dependence of the binodals on the
asymmetry $\eps$ is apparent in both cases. It results from the
explicit dependence of $g_0(\rho)$ on $\eps$ established in
Eq.~\eqref{eq:paramMF}. This dependence probably explains why
run-and-tumble particles hopping on lattices with excluded-volume
interactions~\cite{Thompson:2011:JSM} are not well described by the
coarse-grained theory proposed so far for QSAPs which did not account
for any asymmetric sensing~\cite{Tailleur:2008:PRL}. We can see that
our theoretical predictions are more accurate for small $\eps$, as
expected from the derivation of the hydrodynamic equation given
in~\ref{app:drif-diffusion}.

Our theoretical predictions for the phase diagram of QSAPs rely on
two different approximations.  First, we use a mean-field
approximation to derive the specific expression~\eqref{eq:paramMF} for
$g[\rho]$. For our choice of $v(\rho)$, MIPS occurs only at large
densities so that this approximation works very well except in the small and numerically unresolved Ginzburg interval close to the critical point. Second, our general
theory disregards higher order gradient terms in
Eq.~\eqref{eq:dynamics-general}. This probably explains why the
hydrodynamic description works best fairly close to the critical
point, where interfaces are smoothest and the gradient expansion,
Eq.~\eqref{eq:expansion-v}, most accurate. The quantitative
limitations of our gradient expansion highlights that {\em gradient
  terms directly influence the coexisting densities} through
Eq.~\eqref{eqR}, unlike the equilibrium case.

In addition to giving quantitative predictions for the phase diagrams, our approach
sheds light on the observed universality of the MIPS in QSAPs. For
example, the phase diagram does not depend on the exact shape of the
kernel $K$, which enters Eq.~\eqref{eq:paramMF} through $\ell^2$
which then cancels in the nonlinear transform $R(\rho)$.  Similarly,
Fig.~\ref{fig:vrho} also shows lattice simulations of QSAPs in $1d$
where complete phase separation is replaced by alternating domains (with
densities given by the predicted binodal values). This confirms the
equivalence of continuous (ABP) and discrete (RTP) angular relaxation
dynamics for QSAPs~\cite{Cates:2013:EPL,Solon:2015:EPJST}. Our results, however, 
also expose sensitivity to other microscopic parameters
such as the fore-aft asymmetry $\eps$ which enters $g_0$ and therefore
affects the binodals. This might explain the different
collective behaviors seen in swarms of robots that adapt their speeds
to the density sampled in either the forward or the backward
direction~\cite{Mite:PRX:2016}.

\subsection{Finite-size corrections.}

\begin{figure}
\includegraphics[width=0.45\columnwidth]{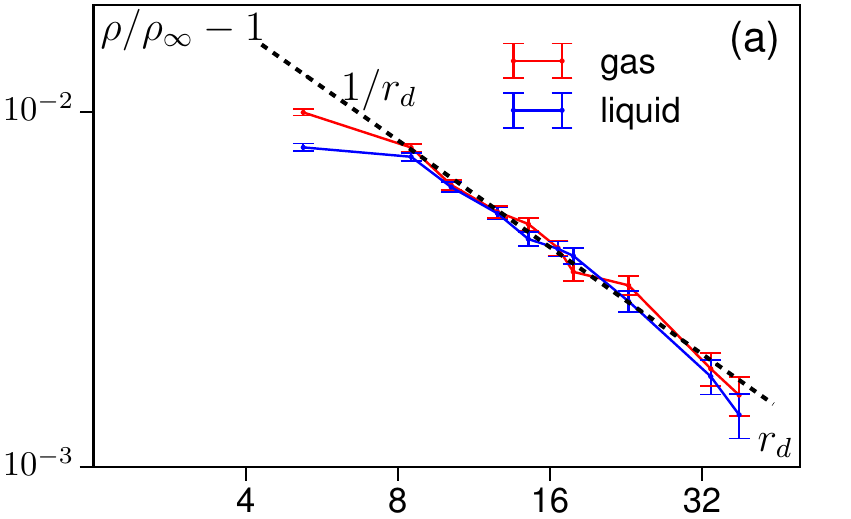}\hspace{0.2cm}
\includegraphics[width=0.45\columnwidth]{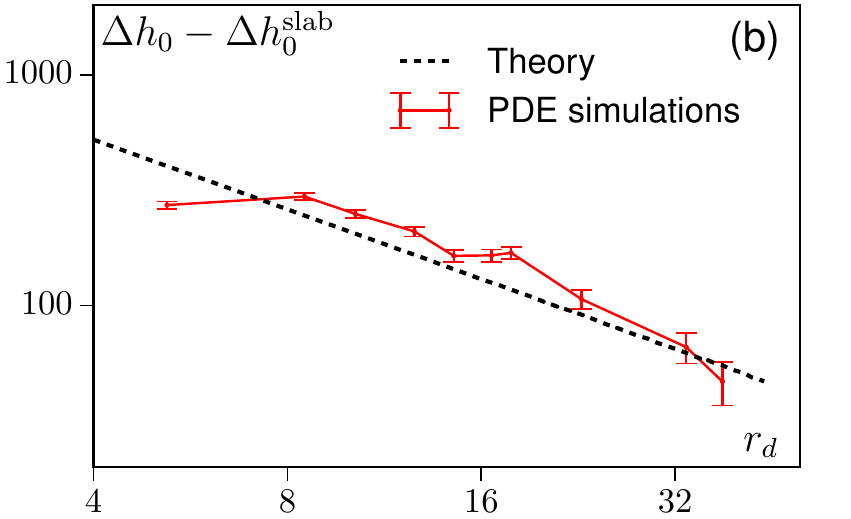}
\caption{Finite size effects in phase-separated QSAPs, measured in 2d
  off-lattice simulations of QSAPs with symmetric sensing
  ($\eps=0$). {\bf (a):} Correction to the coexisting densities. The
  dashed line shows a scaling proportional to $1/r_d$. 
  {\bf (b):} Correction to the generalized pressure $h_0$. The dashed line is the
  predicted leading order behavior $\gamma/r_d$ with the effective
  surface tension $\gamma$ measured independently across a straight
  interface using Eq.~(\ref{eq:surftensgen}). $\Delta h_0^{\rm slab}$
  accounts for the small pressure jump due to Eq.~(\ref{eq:paramMF})
  not being exact, as described in the text. The simulation parameters are $\rho_m=50$,
  $v_0=30$, $v_1=5$, $\tau=1$ and a system size $100\times 100$.}
  \label{fig:vofrho-laplace}
\end{figure}

Similar to the scalar active model of Section \ref{sec:PDEexample}, we
expect finite size corrections when a liquid droplet is formed in a finite system: for a droplet of radius $r_d$, we
expect to leading order in the droplet radius an effective pressure jump across the interface~\eqref{eq:pressurejump}:
\begin{equation}
\Delta h_0(r_d) \simeq \frac 1 {r_d} \int^{r_g}_{r_\ell} \frac{\kappa}{R'}(\partial_r R)^2 dr \;.\label{eq:FSCQSAPs}
\end{equation}
Accordingly, one expect the finite-size corrections to the co-existing densities to decay as $\propto 1/r_d$. In Fig.~\ref{fig:vofrho-laplace}a, we show that the measured binodals
indeed converge towards their asymptotic values in a manner consistent with a $1/r_{d}$ decay.

A quantitative check of~\eqref{eq:FSCQSAPs} is difficult since, first, our derivation of $h_0(R(\rho))$ is based on a number of approximations, and, second, our numerical measurements of the binodals are necessarily noisy. To proceed, we measure $\rho_g(r_d)$ and $\rho_\ell(r_d)$ and construct
$\Delta h_0(r_d) = h_0(\rho_\ell(r_d))-h_0(\rho_g(r_d))$. $\Delta h_0(r_d)$ does not vanish exactly in the large system size limit, nor in a slab geometry in which we measure a small correction
$\Delta h_0^{\rm slab}/h_0(\rho_\ell) \approx 0.1\%$. This systematic
error can stem from several origins, from the gradient expansion to
the mean-field approximation, through limitations in the numerical
accuracy of the density measurement. Though very
small, this error becomes comparable to the Laplace pressure jump for
radii $r_d\apprge 20$, highlighting the numerical challenges in 
measuring these finite-size effects. Nevertheless we show in Fig.~\ref{fig:vofrho-laplace}b that $\Delta h_0(r_d) - \Delta h_0^{\rm slab}$ converges to its asymptotic value
consistently with a $1/r_d$ decay. Furthermore, the prediction of
Eq.~\eqref{eq:FSCQSAPs} can be checked by measuring the prefactor 
$\gamma\equiv \int_{r_g}^{r_\ell} \frac{\kappa}{R'}(\partial_r R)^2
dr$
of this decay in a slab geometry. The corresponding prediction is shown as a dashed
line in Fig.~\ref{fig:vofrho-laplace}b and agrees semi-quantitatively with our numerical results, without any fitting parameters.

To understand why we observe a quasi-quantitative agreement despite relatively small values of $r_d$, it is useful to explicitly expand Eq.~\eqref{eq:halas} as
\begin{equation}
\Delta h_0\simeq \frac{1}{r_d}\int_{r_\ell}^{r_g} dr R' \kappa (\partial_r\rho)^2\left(1-\frac{r-r_d}{r_d}+\frac{(r-r_d)^2}{r_d^2} \right)\;.
\end{equation}
The first order correction to $\Delta h_0 = \frac{\gamma}{r_d}$ is thus given by
\begin{equation}
\Delta h_0 -\frac{\gamma}{r_d} =  - \frac 1 {r_d^2} \int_{r_\ell}^{r_g} dr R' \kappa (\partial_r\rho)^2 (r-r_d).
\end{equation}
Using that, from the definition~\eqref{eqR}, $(R'\kappa)'=-2\lambda R'$, the prefactor $R'\kappa$ can then be expanded around $r=r_d$ as 
\begin{equation}
R'[\rho(r)]\kappa[\rho(r)]\simeq R'[\rho(r_d)]\kappa[\rho(r_d)]-2 \partial_r \rho(r_d)(r-r_d) \lambda[\rho(r_d)] \kappa'[\rho(r_d)].
\end{equation}
Since $\lambda=0$ for QSAPs, we are left with
\begin{equation}
\Delta h_0 -\frac{\gamma}{r_d} =  - \frac {R'[\rho(r_d)] \kappa[\rho(r_d)]} {r_d^2} \int_{r_\ell}^{r_g} dr (\partial_r\rho)^2 (r-r_d),
\end{equation}
which is zero for a symmetric interface so that in that case
\begin{equation}
\Delta h_0 = \frac{\gamma}{r_d} \left[1+{\cal O}\left(\frac 1 {r_d^2}\right)\right].
\end{equation}
For our choice of $v(\rho)$, the density profile is indeed very close to a hyperbolic tangent (data not shown) and the lack of first order corrections for such profiles probably explains the semiquantitative agreement of our numerical results with the $1/r_d$ behaviour.

\section{PFAPs.}
\label{sec:PFAPs}
We now consider the case of self-propelled particles interacting via
pairwise forces, which has attracted considerable interest over the
past few
years~\cite{Fily:2012:PRL,Redner:2013:PRL,Stenhammar:2014:SM,Wysocki:2014:EPL,Yang:2014:SM,Takatori:2015:PRE,Solon:2015:PRL,levis2017active}. We
define the model in Section~\ref{sec:PFAPsmodel} and construct its
hydrodynamic description in Section~\ref{sec:PFAPshydro}. Contrary to
QSAPs, there is no available method to derive accurate
estimates of the coefficients $\lambda(\rho)$ and $\kappa(\rho)$ or to rule out the existence of other terms~\cite{Tjhung:2018:arxiv}. We
discuss in Section~\ref{sec:PFAPsPD} how we can nevertheless follow the path laid out using our generalized thermodynamic formalism to understand how coexistence densities are selected in PFAPs. Finally, finite-size effects are considered in Section~\ref{sec:PFAP-FS}.

\subsection{Model.}
\label{sec:PFAPsmodel} We consider $N$ self-propelled particles in two
dimensions interacting via the repulsive, pairwise additive,
Weeks-Chandler-Andersen potential:
\begin{equation}
  V(r) = 4\epsilon \!\left[ \left(\frac{\sigma}{r}\right)^{12} - \left(\frac{\sigma}{r}\right)^{6} \right]\! + \epsilon 
  \label{WCA}
\end{equation}
with an upper cut-off at $r = 2^{1/6}\sigma$, beyond which $V =
0$. Here $\sigma$ defines the particle diameter, $\epsilon$ determines
the interaction strength, and $r$ is the center-to-center separation
between two particles. Particle $i$ evolves in two dimensions, with
periodic boundary conditions, according to the  Langevin equations:
\begin{equation}
  \dot{\bfr}_i=-\mu \sum_j \grad_i V(|\bfr_i-\bfr_j|)+ \sqrt{2D_t} {\boldsymbol \xi}_i+v_0 \bfu_i;\qquad \dot \theta_i=\sqrt{2D_r}\eta_i.
\end{equation}
Here, $\bfu_i=(\cos\theta_i,\sin\theta_i)$ indicates the direction
of self-propulsion and $\eta_i,{\boldsymbol \xi}_i$ are unit Gaussian
white noises. For simplicity, we only include continuous rotational
diffusion but our results are expected to extend to run-and-tumble
dynamics since these two types of orientational noise have been shown
to lead to the same phase diagram~\cite{Solon:2015:EPJST}.

The full phenomenology of this model requires scanning a
three-parameter phase diagram, parametrized for instance by the Péclet
number $\Pe=3v_0/(\sigma D_r)$~\footnote{Historically, the Péclet
  number was defined as $\Pe=v_0\sigma/D_t$ with translational
  diffusion $D_t$ and a Brownian rotational diffusion
  $D_r=3 D_t/\sigma^2$. It was latter realized that in simulations of
  PFAPs exhibiting MIPS, the translational diffusion has a negligible
  effect on the phase diagram and could be set to zero. This explains
  the factor $3$ in the definition of $\Pe$, although a dimensionless
  run length $l_r=v_0/(\sigma D_r)$ would seem more natural.}, the
packing fraction $(\pi/4)\sigma^2 \rho$ and the potential stiffness
$\mu \epsilon/(v_0 \sigma)$. Here, we focus on the onset of MIPS as
the Péclet number and the packing fraction are varied, disregarding
the role of the potential
stiffness~\cite{Fily:2012:PRL,Redner:2013:PRL,Bialke:2013:EPL,Levis:2014:PRE}. In
practice, we fix $\epsilon=1,\sigma=1,v_0=24,\mu=1$ and vary $D_r$ and
$\rho$. MIPS then occurs at high enough densities when the run-length
$v_0/D_r$ is much larger than the particle size $\sigma$, namely when
$\Pe$ exceeds a threshold value
${\rm Pe}_c\approx
50$~\cite{Fily:2012:PRL,Redner:2013:PRL,Bialke:2013:EPL,Stenhammar:2013:PRL}.

\subsection{Hydrodynamic description.}
\label{sec:PFAPshydro}
Following~\cite{farrell:2012:PRL,Solon:2015:PRL}, we start from the
exact It\=o-Langevin equation for the microscopic density of particles
$\hat \psi(\bfr,\theta)=\sum_{i=1}^N\delta(\bfr-\bfr_i)\delta(\theta-\theta_i)$ at position $\bfr$ with orientation
$\theta$
\begin{equation}
  \label{eq:ito-P-r-theta}
  \partial_t \hat\psi=-\div\left[ v_0\vec u\hat\psi +\hat{\vec I}^{(\theta)}-D_t\nabla\hat\psi +\sqrt{2D_t\hat\psi}\Gvec\eta\right]
  +\partial_\theta\left[D_r\partial_\theta\hat\psi+\sqrt{2D_r\hat\psi}\xi\right]
\end{equation}
where $\hat{\vec I}^{(\theta)}(\vec r,\theta)=-\int d\vec r'\mu\nabla
V(|\vec r'-\vec r|)\hat\rho(\vec r')\hat\psi(\vec r,\theta)$,
$\hat\rho(\bfr)=\int \hat\psi(\vec r,\theta)d\theta$ is the fluctuating
density, and $\Gvec\eta$ and $\Gvec\xi$ are unit-variance Gaussian
white noises of appropriate dimensionality. Denoting averages over noise realizations by angular brackets we define $\rho(\vec r)=\langle \hat\rho(\vec r)\rangle$, $\vec m(\vec r)=\langle \hat \vec m(\vec
r) \rangle$ and $\mathds{Q}(\bfr)=\langle \hat \mathds{Q}(\bfr) \rangle$. Here $\hat {\vec m}(\vec r) = \int d\theta \vec
u\hat\psi(\vec r,\theta)$, is the orientation vector, $\hat \mathds{Q}(\bfr)=\int
d\theta \left( \bfu : \bfu - \mathds{1}/2\right)\hat\psi(\vec
r,\theta)$, is the nematic tensor, and $\mathds{1}$ the identity matrix.

Integrating Eq.~(\ref{eq:ito-P-r-theta}) over $\theta$ and averaging
over noise realizations, the dynamics of $\rho(\vec r,t)$ reads
\begin{equation}
  \label{eq:density-evolution}
  \partial_t\rho=-\div\vec J;\quad \vec J=v_0\vec m+\vec I^{(0)}-D_t\grad\rho
\end{equation}
with 
\begin{equation}
  \vec I^{(0)}(\vec r)=\left\langle \int d\theta \hat{\vec
    I}^{(\theta)}(\vec r,\theta)\right\rangle = -\int d \vec r' \mu\nabla V(|\vec r-\vec r'|) \langle \hat \rho(\bfr) \hat \rho(\bfr') \rangle\;.
\end{equation}
The dynamics of $\vec m$ is then obtained similarly by multiplying
Eq.~(\ref{eq:ito-P-r-theta}) by ${\bf u}$ and integrating over $\theta$. This yields, with an implied summation over repeated indices, 
\begin{equation}
  \label{eq:m-evolution}
  \partial_t m_\alpha=-\partial_\beta \left[ v_0 \left(\mathds{Q}_{\alpha
      \beta}+\frac{\rho\delta_{\alpha\beta}}2\right) +
    \mathds{I}^{(1)}_{\alpha\beta}-D_t\partial_\beta
    m_\alpha\right]-D_r m_\alpha
\end{equation}
where the last term is obtained by integration by parts and we have
defined
\begin{equation}
  \mathds{I}^{(1)}_{\alpha\beta}=-\int d\vec r' \mu\partial_\beta
  V(|\vec r-\vec r'|) \langle \hat \rho(\vec r') \hat m_\alpha(\vec r)
  \rangle\;.
\end{equation}
We stress that, so far, Eq.~(\ref{eq:ito-P-r-theta})
and~(\ref{eq:m-evolution}) are exact, although they are not closed
since they feature $\mathds{Q}$ and the microscopic correlators in
$\vec I^{(0)}$ and $\mathds{I}^{(1)}$ which depend on higher
moments of $\hat\psi$.

As a first approximation, we use that, contrary to $\rho(\vec r,t)$,
$\vec m$ is a fast mode decaying at a rate $D_r$. On time scales
much larger than $D_r^{-1}$, one can thus assume that $m_\alpha$ relaxes
locally to
\begin{equation}
  \label{eq:m-localeq}
  m_\alpha =-\frac{1}{D_r}\partial_\beta \left[ v_0 \left(\mathds{Q}_{\alpha
      \beta}+\frac{\rho\delta_{\alpha\beta}}2\right) + 
    \mathds{I}^{(1)}_{\alpha\beta}-D_t\partial_\beta
    m_\alpha\right].
\end{equation}
The current in Eq.~(\ref{eq:density-evolution}) is then given by
\begin{equation}
\label{eq:currentPFAPS}
J_\alpha= -\left[ D_t+\frac{v_0^2}{2 D_r} \right] \partial_\alpha
\rho-\frac
{v_0}{D_r} \partial_\beta \mathds{I}^{(1)}_{\alpha\beta}+I^{(0)}_\alpha-\frac{v_0^2}{D_r}\partial_\beta \mathds{Q}_{\alpha\beta}
+\frac{D_t v_0}{D_r}\partial_{\beta\beta} m_\alpha \;.
\end{equation}
Interestingly, Eq.~\eqref{eq:currentPFAPS} can be rewritten as
the divergence of a stress tensor
\begin{equation}
J_\alpha = \mu \partial_\beta \sigma_{\alpha\beta}
\label{eq:dynPFAPS}
\end{equation}
with
\begin{equation}\label{eq:sigmaPFAPs}
  \sigma_{\alpha\beta} = - \left[\frac{D_t}{\mu}+\frac{v_0^2}{2\mu D_r}\right]\rho\delta_{\alpha\beta}-\frac{v_0 }{\mu D_r}\mathds{I}^{(1)}_{\alpha\beta}+\sigma^{\rm IK}_{\alpha\beta} -\frac{v_0^2}{\mu D_r} \mathds{Q}_{\alpha\beta}
  +\frac{D_t v_0}{\mu D_r}\partial_\beta m_\alpha
\end{equation}
where we have followed Irving and
Kirkwood~\cite{irving_statistical_1950} (and
Ref.\cite{Kruger2017} in a similar context) and write $I^{(0)}_\alpha =  \mu \partial_\beta \sigma^{\rm
  IK}_{\alpha\beta}$ with 
\begin{equation}\label{eq:sigmaIK}
  \sigma^{\rm IK}_{\alpha\beta}(\bfr)=\frac 1 2 \int d\bfr' \frac{(\bfr - \bfr')_{\alpha} (\bfr - \bfr')_{\beta}}{|\bfr - \bfr'|} \frac{dV(|\bfr - \bfr'|)}{d|\bfr - \bfr'|}\int_0^1d\lambda\langle \hat\rho(\bfr + (1-\lambda)\bfr')\hat\rho(\bfr-\lambda\bfr')\rangle \;.
\end{equation}
We now turn to relate these results to the formalism derived previously.

\paragraph{Generalized pressure and equation of state.}~\newline

The resulting dynamics for $\rho$, with the current given by
Eq.~(\ref{eq:dynPFAPS}), should be compared to the generalized
Cahn-Hilliard equation of Section~\ref{sec:general} with the current
driven by the generalized stress tensor as in
Eq.~(\ref{eq:current-stress}). We see that PFAPs correspond to the
special case $M/R=\mu$, the microscopic mobility. This has important consequences
for the mechanical interpretation of $\Gvec\sigma$. Indeed, one can
see that imposing an external potential $U$ on the particles leads to
\begin{equation}\label{eq:extpot}
  \label{eq:sigma-U}
  \vec J=\mu\div\Gvec\sigma -\mu\rho \nabla U\;.
\end{equation}
In a flux free steady state, $\vec J=\vec 0$ and Eq.~(\ref{eq:sigma-U})
becomes a force balance. Integrating~\eqref{eq:extpot} from a point in the bulk to infinity shows the normal component of $\Gvec\sigma$ to be equal to 
the total force per unit area exerted on a boundary. Indeed, the normal component of $\Gvec\sigma$ exactly coincides in homogeneous phases with the equation of state (EOS) found previously for the mechanical pressure $P$ of PFAPs~\cite{Solon:2015:PRL}. Generalized and mechanical pressure thus coincide for PFAPs and we note, following Section~\ref{sec:generalize-framework}
\begin{equation}
h \equiv -{ \sigma}_{xx}=\frac{D_t}\mu\rho+P^A(x)+P^D(x)+\frac{v_0^2}{\mu D_r} \mathds{Q}_{xx}-\frac{D_t v_0}{\mu D_r} \partial_x m_x \label{eq:PRPFAPS}
\end{equation}
where we have defined, following earlier notation~\cite{Solon:2015:PRL}, the ``active'' contribution to the pressure $P^A$ and a ``direct'' passive-like part $P^D$:
\begin{equation}
  \label{eq:pressures}
  P^A=\frac{v_0^2}{2\mu D_r}\rho+\frac{v_0}{\mu D_r} \mathds{I}^{(1)}_{xx}; \qquad P^D=-\sigma^{\rm IK}_{xx}.
\end{equation}
Note that $P^A$ is sometimes also called ``swim pressure''~\cite{Brady:2014:PRL}, even though neglecting the pressure of the surrounding fluid to describe the phase separation of actual swimmers is problematic.

The value of the pressure in a homogeneous phase of density $\rho_0$ is then given by
\begin{equation}
h \left[\rho(x)=\rho_0 \right]  \equiv h_0 (\rho_0)= \rho_0 \frac{D_t}{\mu} + P^A_0+P^D_0 \;,
\end{equation}
where $P^A_0$ and $P^D_0$ are the values taken by $P^A$ and $P^D$ in homogeneous disordered phases of density $\rho_0$. This allows us to identify, in analogy with Eq. \eqref{eq:h0h1},
\begin{equation}
h=h_0(\rho(x))+h_1([\rho],x) \;.
\end{equation}
Note that while the structure is similar to Eq. \eqref{eq:h0h1}, there is no gradient expansion taken here -- $h_1$ is exact, formally containing gradients of all orders. Its expression is given by
\begin{equation}\label{eq:h1pfaps}
h_1 =P^A_1[\rho]+P^D_1[\rho]+\frac{v_0^2}{\mu D_r} {\mathds{Q}_{xx}}-\frac{D_t v_0}{\mu D_r} \partial_x m_x.
\end{equation}
where $P^{A/D}_1\equiv P^{A/D}-P^{A/D}_0$ contains the interfacial contributions to the active and the direct pressures. 
The terms in $m_x$ and $\mathds{Q}_{xx}$ are purely interfacial since they
vanish in the (disordered) bulk phases. We now show that the phase equilibria in PFAPs can be understood using these results with the ideas of Section~\ref{sec:general}.  

\subsection{Phase equilibria in PFAPs.}
\label{sec:PFAPsPD}
One way forward would be to construct an explicit gradient expansion for $h$ in terms of $\rho$ and obtain closed expressions for $h_0$ and $h_1$. This would then allow us to find $R(\rho)$ and $\phi(R)$ analytically as was done for QSAPs in Section~\ref{sec:QSAPs}. Despite
the extensive literature on PFAPs, such a gradient
expansion has not yet been presented, but could be accomplished for instance by using a low-density virial approximation. Such a route would possibly lead to qualitative predictions for the phase diagram, but our goal here is to show that our formalism \emph{quantitatively} accounts for the phase equilibria of PFAPs, and we thus do not want to rely on such approximations.  We thus proceed differently, using an approach where we instead \emph{measure} the gradient terms to quantitatively verify the validity of our formalism for PFAPs.  
\begin{figure}
\begin{center}
\includegraphics[width=80mm]{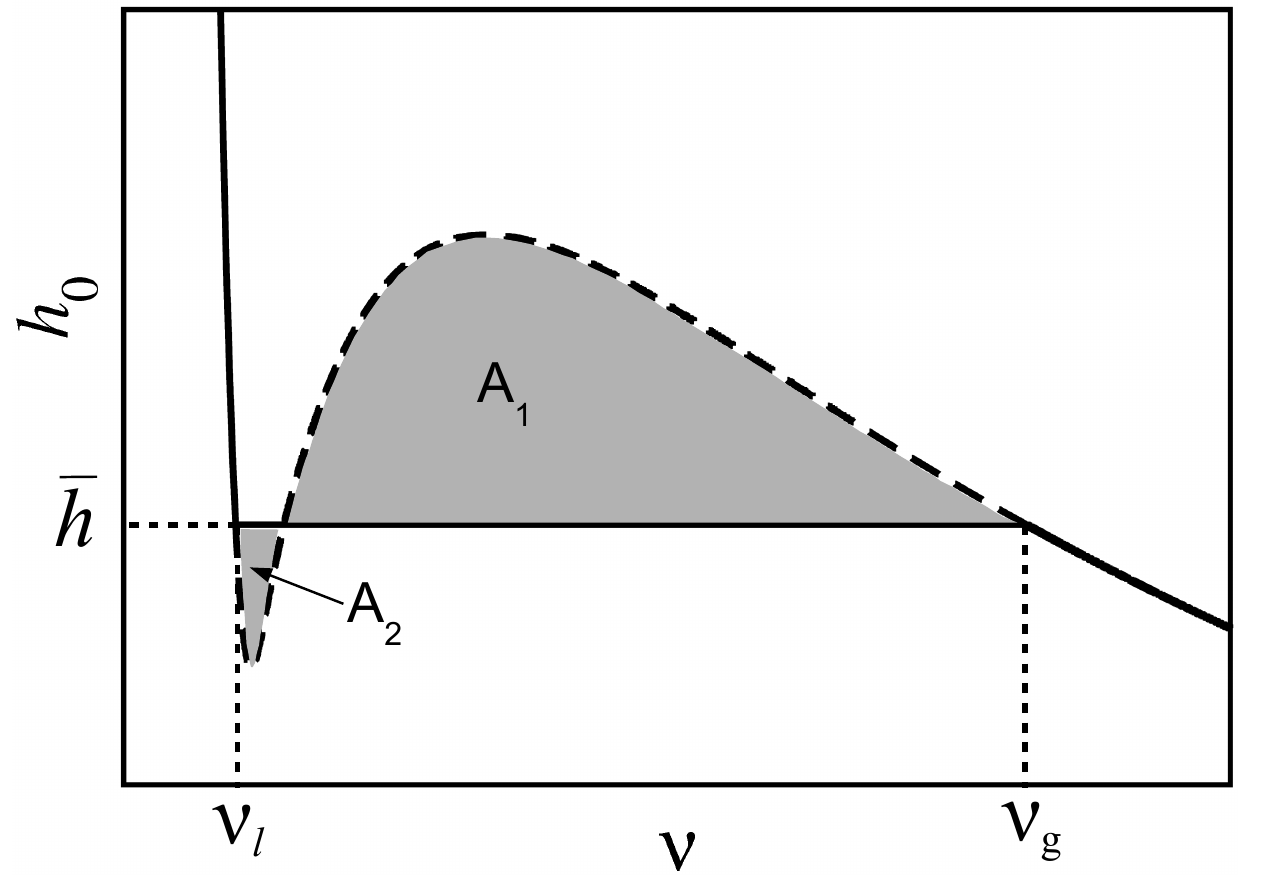}
\end{center}
\caption{Schematic picture of the violation of the Maxwell equal area
  construction. The black non-monotonic line shows the equation of
  state $h_0(\nu)$, dashed in the part where homogeneous systems are
  unstable with respect to phase separation.  The violation of the Maxwell equal-area construction is
  quantified by $\Delta A=A_1-A_2\neq 0$.}\label{Maxwell}
\end{figure}

As with the other systems, we first consider the case of a
{macroscopically} phase separated system, for which the liquid-gas interface is locally flat and perpendicular to $\mathbf{\hat{x}}$. As in Section~\ref{sec:binodals}, in a flux-free steady state, $h=\bar h$ is constant across the interface so that the pressure is equal in coexisting phases
\begin{equation}
  \label{eq:eq-pressure-PFAPs}
  h_0(\rho_g)=h_0(\rho_\ell)=\bar h.
\end{equation}
To construct the phase diagram, we need to complement this equality by a second constraint. Since we do not have any closed expression for the interfacial terms $h_1$, we cannot use a Maxwell construction in the $(h_0,\upupsilon=R^{-1})$ plane as was done in Section~\ref{sec:binodals}. Instead, we measure the {\emph{violation}} of the {\emph{equilibrium}} Maxwell construction in the $(h_0,\nu \equiv \rho^{-1})$ plane, {schematically} depicted in Fig.~\ref{Maxwell}, with $\nu=1/\rho$ {the free volume per particle}:
\begin{eqnarray}
  \label{eq:ABPs-unequal-area}
  \int_{\nu_\ell}^{\nu_g}(h_0(\nu)-\bar h)d\nu= \int_{x_g}^{x_\ell}h_{1}\,\partial_x \nu\, dx\equiv \Delta A\;.
\end{eqnarray}
Here $h_0(\nu)$ is the pressure-volume EOS, $\bar h$ is the pressure of coexisting phases and  $\Delta A\neq 0$  {\em directly} quantifies the violation of the Maxwell construction for PFAPs~\cite{Solon:2015:PRL}. 

Given the value of $\Delta A$, Eqs~\eqref{eq:eq-pressure-PFAPs} and~\eqref{eq:ABPs-unequal-area} are two independent constraints satisfied by $\rho_\ell$ and $\rho_g$.  A fully predictive theory would thus evaluate $\Delta A$ analytically and then solve~\eqref{eq:eq-pressure-PFAPs} and~\eqref{eq:ABPs-unequal-area} to obtain the values of the binodals and the coexisting pressure $\bar h$. Here, instead, we use a numerical measurement of $\Delta A$ to construct the phase diagram. Although less predictive than knowing $h_0$ and $h_1$ analytically, our method clearly illustrates how the violation of the Maxwell construction, due to the role played by the interfaces, selects the binodals.

\begin{figure}
  \begin{center}
\includegraphics[width=0.9\columnwidth]{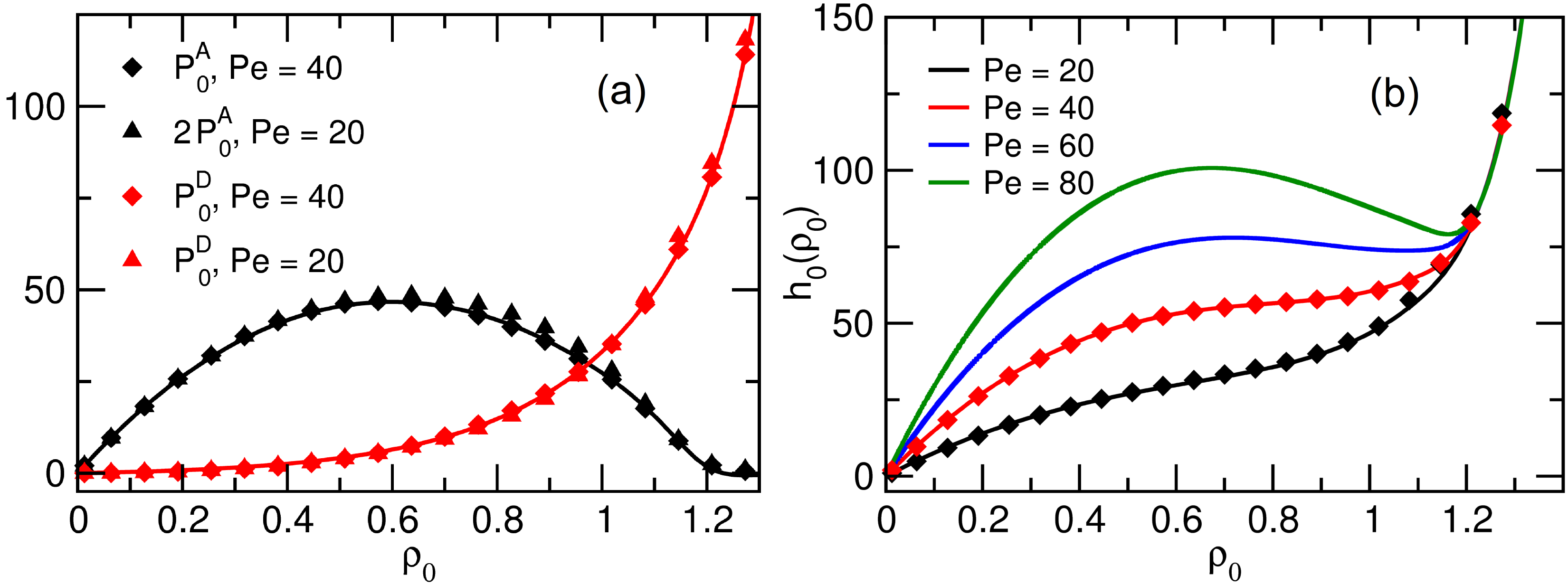}
\caption{Construction of the equation of state for the pressure of
      PFAPs. {\bf (a)} $P^A$ and $P^D$ (symbols) as a function of the average
      density $\rho_0$ measured in simulations of homogeneous systems
      ($\mathrm{Pe} < \mathrm{Pe}_c$). The solid lines show
      fits to the data at $\Pe = 40$ using the functional forms $P_0^A$ and $P_0^D$ detailed
      in~\ref{app:EOS}. As seen from the rescaling, $P^A_0$
      scales linearly with $\Pe$ while $P^D_0$ is independent of
      $\Pe$. We use this scaling to extrapolate the equation of state
      to the region where the system phase separates ($\mathrm{Pe} >
      \mathrm{Pe}_c$). {\bf (b)} The full equation of state (solid lines) for the
      pressure $h_0 = P_0^A + P_0^D$, with symbols denoting numerical
      measurements. {The curves for $\Pe=60$ and $\Pe=80$ are extrapolated 
      from measurements at $\Pe=40$ using the above scaling arguments.}}\label{EOS}
  \end{center}
\end{figure}

\paragraph{Numerical strategy and results}~\newline To numerically
construct the phase diagram, we first derive an approximation to the
bulk equation of state $h_0(\rho)$. Then, we measure $h(x)$
numerically via Eq.~\eqref{eq:PRPFAPS} from which we subtract
$h_0(\rho(x))$ to obtain $h_1$, which is integrated to obtain the
numerical value of $\Delta A$. The right hand side of
Eq.~\eqref{eq:ABPs-unequal-area} is then held constant at this value,
and the binodals are determined as the intersect between the EOS
$h_0(\nu)$ and a horizontal line of ordinate $\bar h$ whose value is
adjusted until it satisfies Eq.~\eqref{eq:ABPs-unequal-area}. Note
that, for the parameter range of interest here, the two contributions
to $h$ proportional to $D_t$ are negligible and we thus discard them
hereafter.

\begin{figure}
  \centering
\includegraphics[width=0.9\columnwidth]{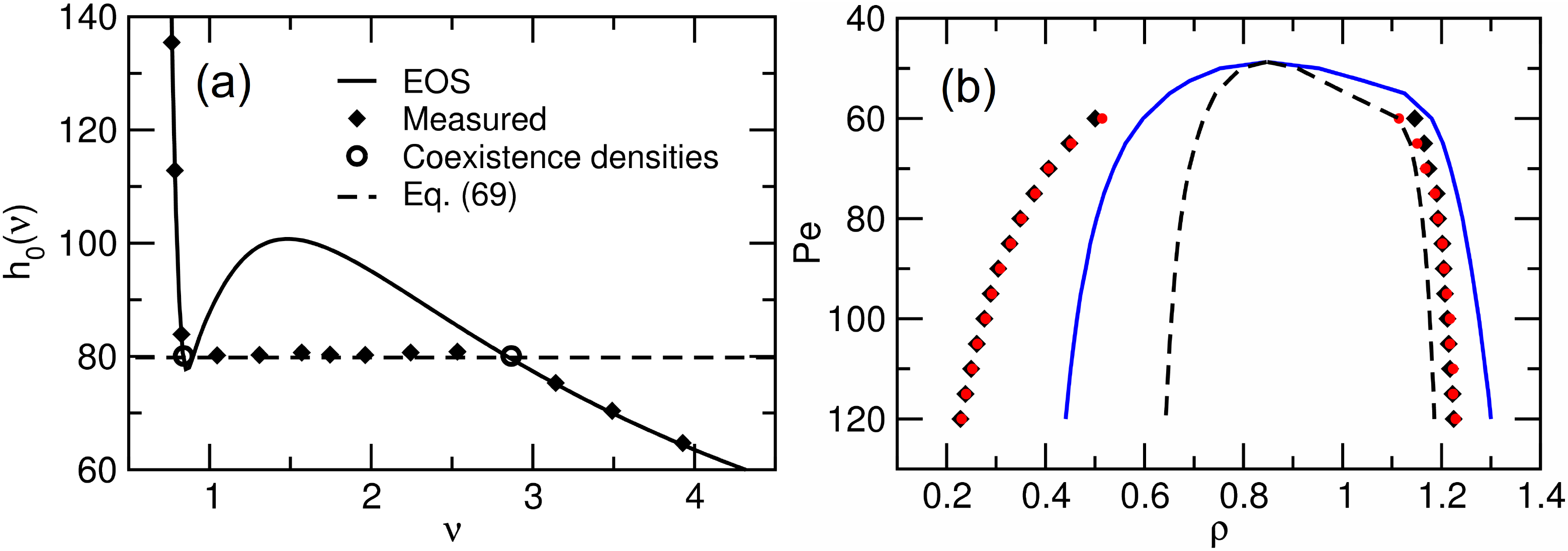}
\caption{{\bf (a)} Unequal-area construction on the equation
    of state $h_0(\nu)$ (solid lines) at Pe = 80, using the value of $\Delta A$ obtained by measuring the gradient terms. Open circles correspond to the measured binodals, and filled diamonds correspond to the pressures measured in the numerics. Note that the generalized pressure remains constant along the tie line. Dashed lines indicate pressures predicted by the unequal-area construction across the tie line.
    {\bf (b)} Phase diagrams of PFAPs,
    measured numerically (diamonds), through our prediction
    Eq.~(\ref{eq:ABPs-unequal-area}) (red circles), and from the equilibrium Maxwell
    construction with $\Delta A=0$ (blue line). The dashed line corresponds to the
    boundaries of the spinodal region $h_0'(\rho) < 0$.}
  \label{fig:ABPs}
\end{figure}

\begin{enumerate}
\item We first construct an analytical approximation for the pressure
  $h_0(\rho_0)$ by measuring the active and direct pressures from an
  ABP simulation in the \textit{homogeneous} region
  ($\Pe < \Pe_{\rm c}$) using Eqs.~\eqref{eq:pressures}. Following the
  route proposed in~\cite{Solon:2015:PRL}: we then apply scaling
  arguments to extrapolate the EOS into the two-phase region
  $\Pe>\Pe_c$.  Fig.~\ref{EOS} {explains and verifies the proposed
    scaling} in the low-Péclet region and shows the resulting EOS for
  $P^A_0$ and $P^D_0$. Details about the numerical procedure (refined
  with respect to Ref.~\cite{Solon:2015:PRL}) can be found
  in~\ref{app:EOS}.

\item The next step is to numerically determine $h_1$ using
  Eq.~\eqref{eq:h1pfaps} and, through it, the value of $\Delta A$. In
  numerical simulations of phase-separated systems in a slab geometry
  (see Fig. \ref{pressures}a), we thus measure the profiles $\rho(x)$,
  $P^A(x)$, $P^D(x)$ and $\mathds{Q}_{xx}(x)$ across the interface
  (see Fig. \ref{pressures}b-c). Using the EOS $P^A_0(\rho)$ and
  $P^D_0(\rho)$ from (i) together with the measured density profile
  $\rho(x)$ we obtain the gradient contributions to the active and
  direct pressures as $P_1^{A/D}(x) = P^{A/D}(x) - P^{A/D}_0(\rho(x))$
  (Fig. \ref{pressures}d). Together with $\mathds{Q}_{xx}(x)$, this
  directly provides $h_1(x)$ and hence the value of $\Delta A$ in
  Eq.~(\ref{eq:ABPs-unequal-area}).

\begin{figure}
\begin{center}
\includegraphics[width=0.9\textwidth]{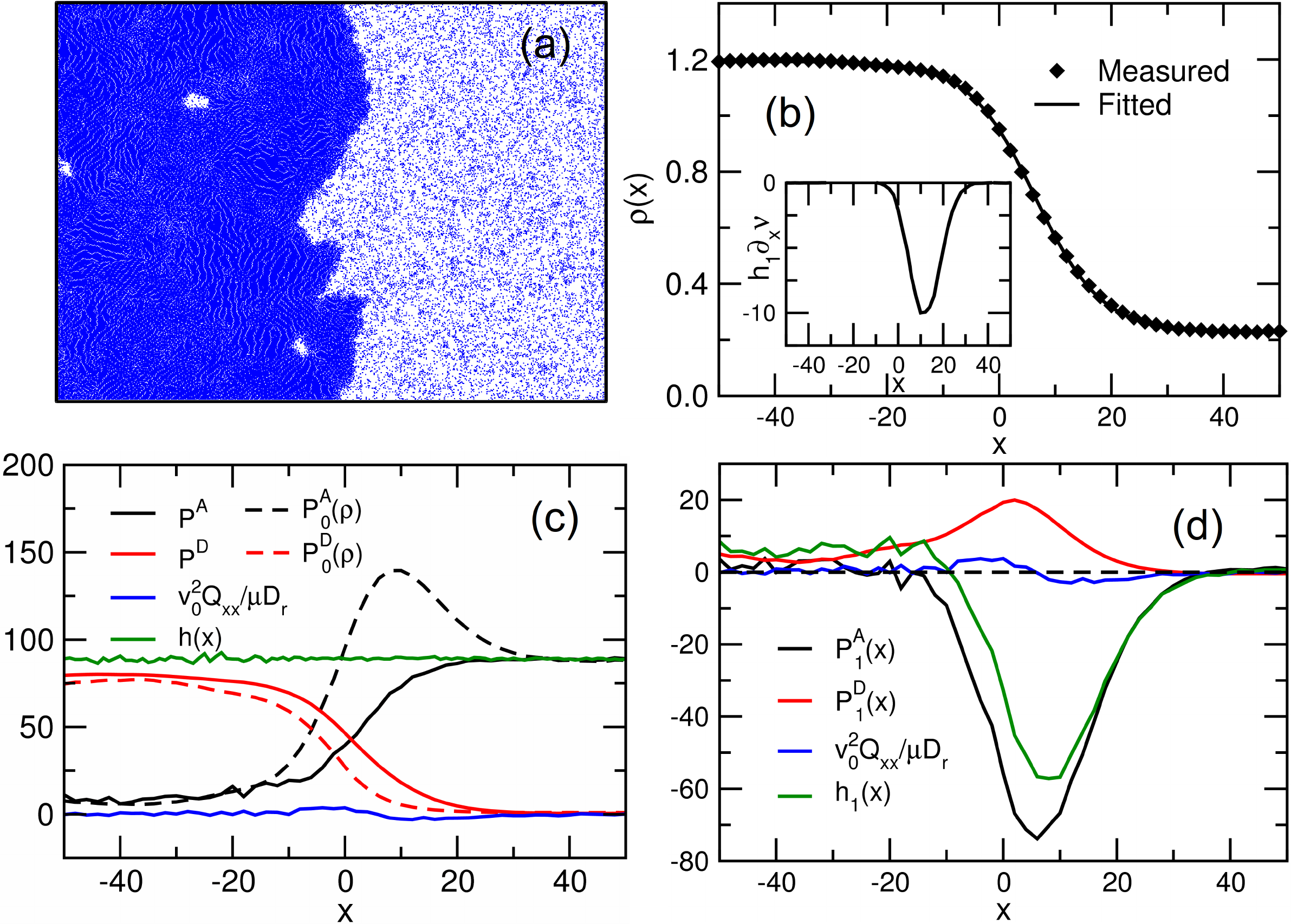}
\caption{{\bf (a)} Close-up of a snapshot showing the interfacial
  region in a phase-separated system at $\Pe = 120$. {\bf (b)} Density
  field $\rho(x)$ across the interface in (a), averaged over $t$ and
  $y$. The solid line is a fit to a hyperbolic tangent
  function. Inset: Plot of $h_1 \partial_x \nu$ across the
  interface. The area under the curve quantifies the violation
  $\Delta A$ of the Maxwell
  construction~\eqref{eq:ABPs-unequal-area}. {\bf (c)} Profiles of the
  total pressure $h(x)$ and its three non-negligible components $P^A$,
  $P^D$ and $v_0^2\mathds{Q}_{xx}/\mu D_r$ (solid lines). The dashed
  lines correspond to the local contributions $P^A_0(\rho(x))$ and
  $P^D_0(\rho(x))$ that are predicted by the equation of state for a
  homogeneous system at density $\rho(x)$. {\bf (d)} The interfacial
  contributions to the pressure, entering $h_1$ in
  Eq.~\eqref{eq:h1pfaps}.} \label{pressures}
\end{center}
\end{figure}

\item Using the equation of
  state $h_0(\nu)$, we now adjust $\bar h$ in
  Eq.~\eqref{eq:ABPs-unequal-area} until $\Delta A$ matches the value
  computed in step (ii) as shown in Fig.~\ref{fig:ABPs}a. The resulting $\bar h$ and the corresponding
  two values of $\nu_g$ and $\nu_\ell$ constitute our prediction for
  the pressure at coexistence and the binodals.
\end{enumerate}

As seen in Fig.~\ref{fig:ABPs}b, the predicted coexistence densities match very well
the measured ones. We stress again that this is not a first principle
prediction, since we do not use an analytic expression for the gradient terms, 
which thus have to be measured numerically. Nevertheless, the
excellent agreement confirms the scenario proposed in
Section~\ref{sec:general} for MIPS: {unlike in equilbrium}, 
the interfacial contributions are
essential in fixing the coexistence densities. Indeed, 
the equilibrium Maxwell construction (equivalent to taking $\Delta
A=0$ in Eq.~(\ref{eq:ABPs-unequal-area})) clearly fails to account for the
phase diagram of PFAPs, as shown in
Fig.~\ref{fig:ABPs}b. Therefore, the interfacial contributions have
to be accounted for, either by defining an effective density as in
Sections~\ref{sec:general}, \ref{sec:PDEexample} and \ref{sec:QSAPs}, or
by quantifying the violation of the equilibrium constructions, as demonstrated
here. 

We finally note that the behavior of the interfacial terms $P^A_1$ and $P^D_1$ in Fig.~\ref{pressures}d  can be qualitatively understood as arising from the polarization of the gas-liquid interface. Since a particle at the interface is on average oriented \emph{towards} the (denser) liquid phase, \emph{i.e.}, up the density gradient, it experiences a more efficient collisional slow-down than it would in an isotropic environment at the same local density. Since $P^A[\rho]$ is proportional to the effective swim-speed 
$v$, this will yield a \emph{lower} $P^A$ than in the isotropic phase, and thus a
negative $P_1^A$. Conversely, as $P^D$ is proportional to the amount of 
repulsive particle contacts experienced by the particle, the same 
argument will lead to a positive interfacial contribution $P^D_1$ to the direct pressure, 
confirming the observations in Fig.~\ref{pressures}d. Since these two terms give the 
dominant contributions to $h_1$, we thus conclude that, at the microscopic level, 
the phase coexistence densities in PFAPs is controlled by the polar ordering of 
particles at the gas-liquid interface.

\subsection{Finite-size corrections.}
\label{sec:PFAP-FS}

We now turn to study the finite-size corrections to the phase
equilibria of PFAPs. {As previously, we consider a circular droplet of
radius $r_d$ (see Fig.~\ref{fig:scheme})}. Following
Section~\ref{sec:FFS-GT}, the pressure jump across the interface is given
at leading order in $1/r_d$ by
\begin{equation}
\Delta h_0= \frac \gamma {r_d};\qquad \gamma=\int_{x_\ell}^{x_g} (\sigma_{yy}-\sigma_{xx}) dx,
\label{eq:PFAPSDh}
\end{equation}
where the surface tension $\gamma$ is measured across a planar interface
perpendicular to $\mathbf{\hat{x}}$. We follow the
same route as for QSAPs and measure independently $\Delta h_0$ and
$\gamma$ in numerical simulations to characterize the finite-size corrections to the phase coexistence.

\begin{figure}
\begin{center}
\includegraphics[width=.5\textwidth]{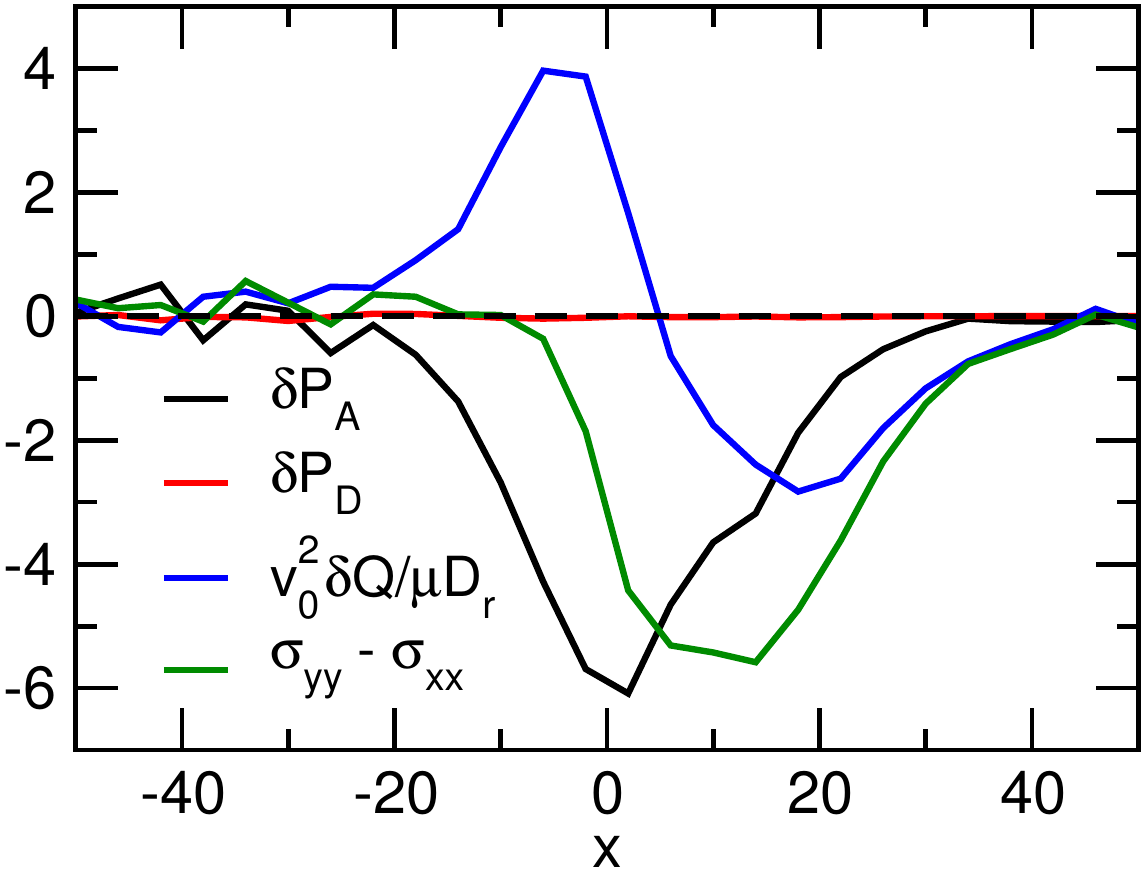}
\caption{The three contributions to the difference between the
  tangential and normal stress components as defined in
  Eqs. \eqref{eq:DeltaPA}--\eqref{eq:DeltaQ}, measured from a
  simulation in slab geometry at $\Pe=100$. The integral of the total
  stress difference $\sigma_{yy} - \sigma_{xx}$ across the interface
  is the effective ``surface tension'' $\gamma$ defined in
  Eq.~(\ref{eq:PFAPSDh}).}\label{fig:gammaterms}
\end{center}
\end{figure}

To understand the different contributions to $\gamma$, we introduce the difference between the $xx$ and $yy$ components for each term in the stress tensor~(\ref{eq:sigmaPFAPs})
(recall that the terms proportional to $D_t$ are negligible):
\begin{eqnarray}
\delta P_A&=&\frac{v_0}{\mu D_r} (\mathds{I}_{xx}^{(1)}-\mathds{I}_{yy}^{(1)}) \label{eq:DeltaPA} \\
\delta P_D&=&-\sigma^{\rm IK}_{xx} +\sigma^{\rm IK}_{yy}\label{eq:DeltaPD} \\
{\delta \mathds{Q}}&=& \mathds{Q}_{xx}-\mathds{Q}_{yy}\;. \label{eq:DeltaQ} 
\end{eqnarray}
The surface tension is then given by
\begin{equation}
\gamma=\int_{x_\ell}^{x_g} \left( \delta P_A+\delta P_D+\frac{v_0^2}{\mu D_r} \delta \mathds{Q} \right) dx,
\label{eq:gamma-contributions}
\end{equation}
These three contributions and their sum, $\sigma_{yy}-\sigma_{xx}$,
are plotted in Fig.~\ref{fig:gammaterms}, as measured across a
flat (on average) interface; the resulting integral yields the estimate
$\gamma \approx -140$ in the units of our simulations. Interestingly, the contribution of the
direct term $\delta P_D$ is completely negligible, in contrast to the equilibrium case in which the phase
separation is due to attractive forces, which also determine the surface
tension. Here, the main contributions stem from the anisotropy of the
active pressure in the interface, as well as from the anisotropic
nematic order of the particles in the interfacial region.
We furthermore note that the resulting value of the surface tension is
\emph{negative}, which confirms the finding of
Ref.\cite{Bialke:2015:PRL} and can be rationalized
following~\cite{Lee:2017:SM} by considering the escape angle of an
active particle exiting a curved interface.

\begin{figure}[h]
\begin{center}
\includegraphics[width=160mm]{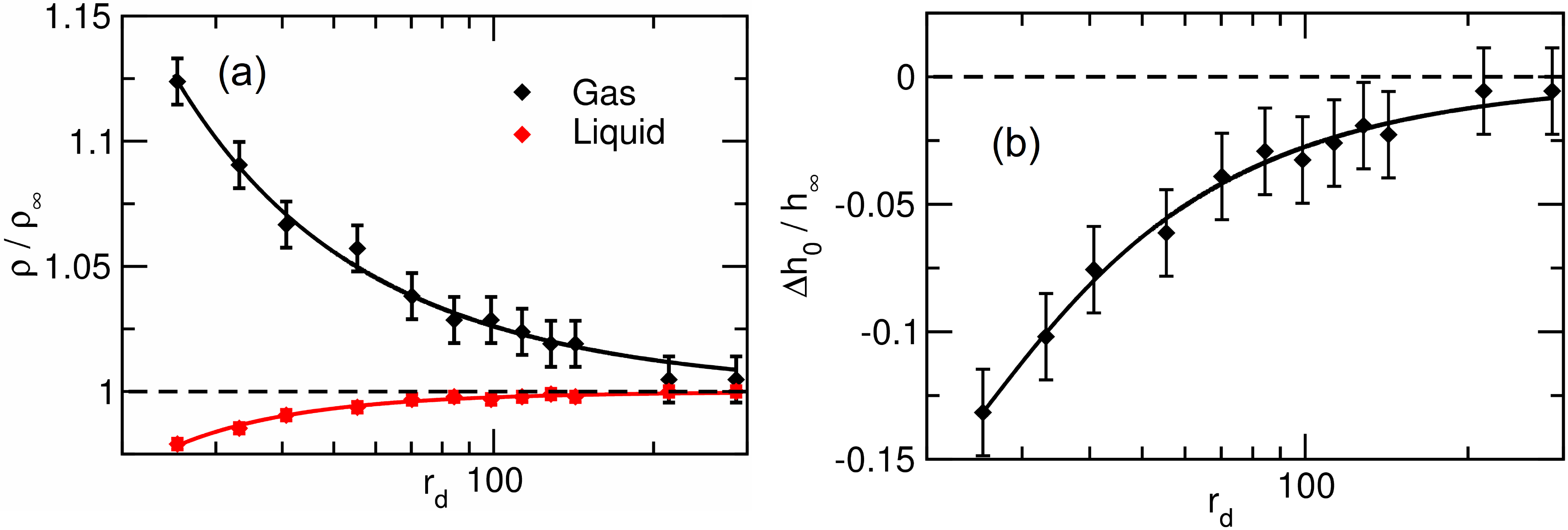}
\caption{{\bf (a)} Coexisting densities measured numerically as a
  function of the droplet radius for $\Pe=100$, normalized with the
  corresponding densities $\rho_{\infty}$ in the slab geometry
  (\emph{i.e.}, $r_d = \infty$). The solid lines
  indicate fits to the measured data using the function
  $\rho/\rho_{\infty} = 1 + c_1 / r_d + c_2 / r_d^2$, with $c_1$ and
  $c_2$ fitting parameters. Note that these measurements are very sensitive to the definition of coexisting densities, e.g. using 
  the positions of the peaks of maximum probability in the
  distribution $P(\rho)$ of local density $\rho$ vs using the average of such peaks, so that we can expect at best semiquantitative agreement with theory. The droplet
  radii $r_d$ are estimated from the phase volumes obtained from the
  integral of the respective peaks in $P(\rho)$.  {\bf (b)} The
  corresponding difference in coexistence pressure $\Delta h_0$,
  obtained from the densities in (a) using the numerical EOS, and
  normalized with the pressure $h_{\infty}$ for a flat
  interface. Solid lines show fits to $\Delta h_0/h_{\infty} = c_1 /
  r_d + c_2 / r_d^2$, where the fitting parameter $c_1 h_{\infty}
  \approx -230$ is an estimation for the surface tension
  $\gamma$.}\label{Laplace}
\end{center}
\end{figure}

We now evaluate the {effective Laplace pressure} $\Delta h_0$ for
curved droplets of different radii. Although this quantity is in
principle directly measurable in simulations, it is numerically
challenging due to the large fluctuations in the local pressure.  We
thus instead proceed similarly as for QSAPs, by first accurately
measuring the coexisting densities in finite systems in which a liquid
droplet of radius $r_d$ coexists with a vapor background. These are
shown in Fig.~\ref{Laplace}a, showing that the liquid phase is
effectively depleted for finite $r_d$, hence confirming the heuristic
argument given in~\cite{Lee:2017:SM}.  The correction to the
coexistence densities is again found to be compatible with a $1/r_d$
decay. The pressure jump can then be computed using the equation of
state and the measured densities as
$\Delta h_0=h_0(\rho_\ell)-h_0(\rho_g)$, shown in
Fig.~\ref{Laplace}b. To extract the leading order behavior in $1/r_d$,
we fit $\Delta h_0(r_d)$ with two parameters, using a function
$c_1/r_d+c_2/r_d^2$. The second-order term is necessary because the
width of the liquid-vapor interface is large ($\approx 40$, see
Fig. \ref{pressures}b) so that the assumption of large $r_d$ does not
hold.  {The leading-order coefficient from the fit in
  Fig.~\ref{Laplace} corresponds to} a value of $\gamma\approx -230$,
to be compared to $\gamma=-140$ measured across the straight interface
in Fig.~\ref{fig:gammaterms}.  The sign and order of magnitude are
thus correctly captured, in spite of the many approximations and
numerical difficulties inherent in these measurements.

We stress that the procedure we detail above retains all the gradient
terms entering $\Gvec \sigma$ through $h_1$, and hence accounts for
the negative value of $\gamma$. As explained before, we have not,
however, carried out explicitly a gradient expansion of
$\Gvec \sigma$. Therefore, we do not know whether PFAPs can be
quantitatively described by Eq.~\eqref{eq:current-stress}
and~\eqref{eq:stress-tensor}. We have shown that, in the
  formalism of Section~\ref{sec:general}, phase-separated solutions
  are compatible with a negative $\gamma$. However, the finite size
  corrections derived in Section~\ref{sec:FFS-GT} are constrained by
  the equality of generalized chemical potential in the two phases
  which imposes that the density correction $\rho-\rho_\infty$ take
  the same sign in the two phases. This is at odds with the
  observation of Fig.~\ref{Laplace}(a), thereby suggesting that PFAPs
  are not fully described by our generalized Cahn-Hilliard equation. A
  promising suggestion is that the finite size effects of PFAPs are
  best described by a more general gradient expansion which would
  imply the analogue of a Laplace pressure jump for the chemical
  potential~\cite{Tjhung:2018:arxiv}.


\section{Change of ensembles}\label{sec:Ensemble}
One powerful aspect of equilibrium thermodynamics is that it relates
the physical states of a system under different environmental
constraints. Beyond its engineering value, the existence of several
ensembles provides useful theoretical tools to study phase
transitions~\cite{Binder:1987:Rep}. Similar developments for
non-equilibrium systems have however proven
difficult~\cite{Bertin:2006:PRL,Bertin:2007:PRE,Dickman:2016:NJP}.
Interestingly, our formalism allows some progress.
 
 \begin{figure}
  \centering
  \includegraphics[width=1.0\columnwidth]{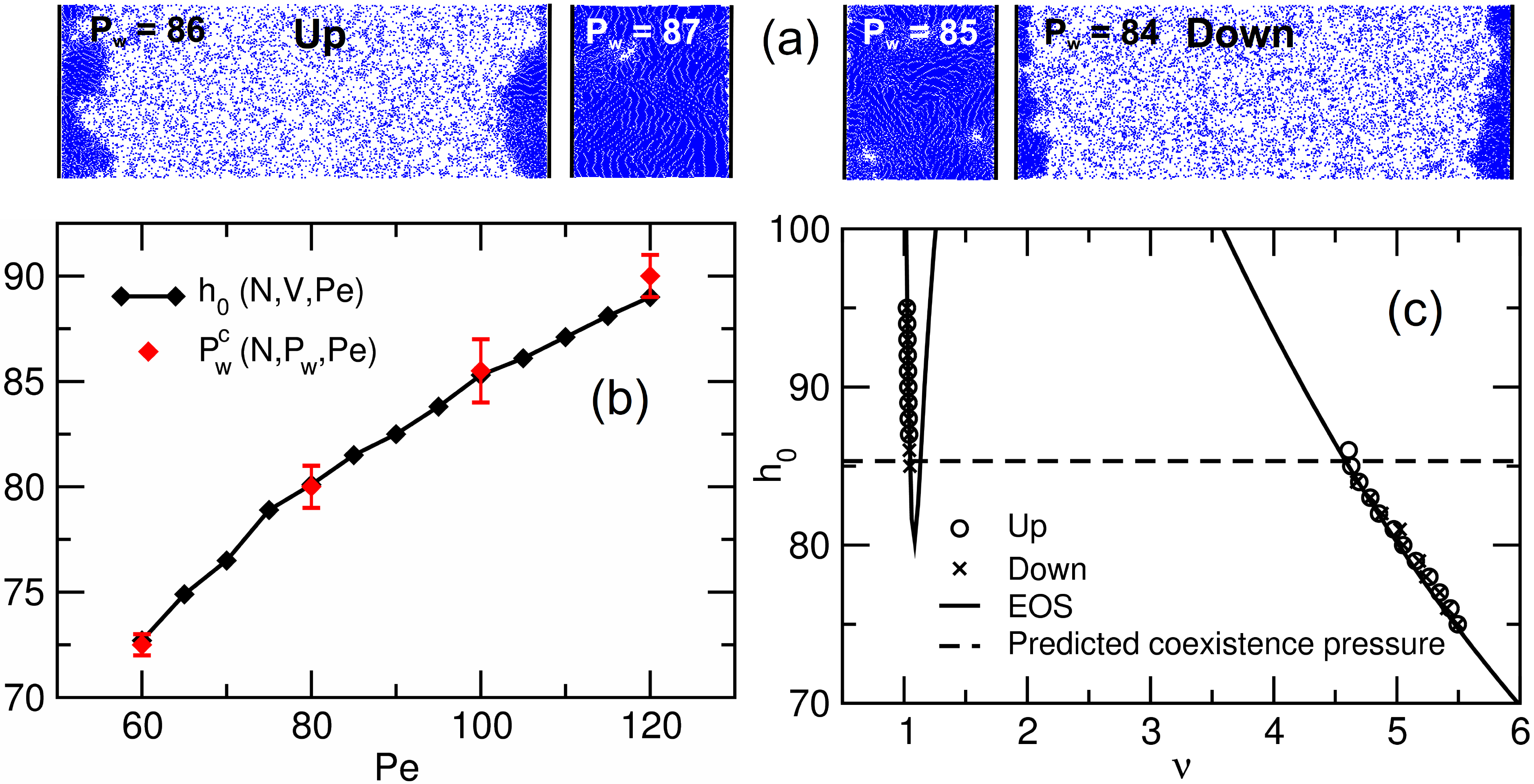}
  \caption{PFAPs in the isobaric $N, P_w, \Pe$ ensemble. {\bf (a)}:
    Snapshots from PFAP simulations with a mobile wall imposing a
    pressure $P_w$ at $\Pe=100$ during a slow upwards (left) and
    downwards (right) pressure ramp (for movies, see \cite{supp}). 
    In the isobaric ensemble, the phase transition becomes discontinuous, 
    in contrast to the phase coexistence observed in constant-volume 
    simulations. {\bf (b)}: For each P\'eclet number, the discontinuous 
    phase transition (red symbols) occurs when the imposed pressure $P_w^c$ 
    equals the mechanical pressure of coexisting gas and liquids in the isochoric
    ensemble (black symbols). {\bf (c)}: When ramping the imposed
    pressure slowly up or down across the transition, the measured
    phase densities (symbols) fall on the pressure equation of state
    (solid black line), with a small hysteresis loop centered around
    the coexistence pressure (horizontal dashed
    line).}\label{fig:NPPe}
\end{figure}

We adapt our previous constant volume (isochoric) simulations to
consider an isobaric (constant pressure) ensemble. PFAPs or QSAPs are
now confined by mobile harmonic walls, subject to a constant force
density $P_w$ which imposes a mechanical pressure $P=P_w$ 
(see Fig. \ref{fig:NPPe}a and movies in \cite{supp}). 
Since $P=h_0$ is a generalized thermodynamic variable for PFAPs, we expect,
as in equilibrium, that the coexistence \textit{region} of the
isochoric ($N,V,\Pe$) ensemble collapses onto a coexistence
\textit{line} in the isobaric ($N,P,\Pe$) case, corresponding to the
pressure at coexistence in the isochoric ensemble (see
Fig~\ref{fig:NPPe}b). Inposed-pressure loops carried out by slowly ramping
up and down $P_w$ then lead to small hysteresis loops around the value
of $P_w$ corresponding to coexistence. These loops would vanish in the
large system size limit for quasi-static ramping of $P_w$ (see
Fig~\ref{fig:NPPe}c).

\begin{figure}
  \begin{center}
    \includegraphics[width=.4\columnwidth]{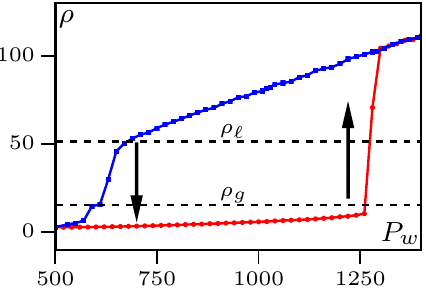}
    \caption{For QSAPs, the volume (or here the density at fixed
      particle number $N=150000$) is not single-valued in the imposed
      mechanical pressure $P_w$, leading to large hysteresis
      loops. Note that the mechanical pressures $P_w$ corresponding to
      liquid and gas binodals are different, as expected. Parameters:
      $\rho_m=25$, $v_0=20$, $v_1=5$, $\tau=1$, vertical size
      $L_y=50$.}\label{fig:NPTQSAPs}
  \end{center}
\end{figure}

In contrast, for QSAPs the mechanical pressure $P$ is unrelated to
either of the generalized variables $g_0, h_0$. The same value of
$P_w$ may thus lead to different states of the system depending on its
history: the Gibbs phase rule does not apply for QSAPs in this
ensemble. This translates into large hysteresis loops when slowly
cycling $P_w$, as shown in Fig~\ref{fig:NPTQSAPs}.

On a fundamental level, the different relationship between
thermodynamical and mechanical observables can be related to the
presence or absence of an effective momentum conservation in the
steady state~\cite{Fily:2017:JPA}. From a more practical point of view,
this can be traced back to the fact that adding an external potential
$U$ to PFAPs gives a simple force balance equation in a flux-free steady state
\begin{equation}
  \label{eq:sigma-U-2}
  \rho {\nabla U}=\div\Gvec\sigma.
\end{equation}
This makes the mechanical pressure a state variable for PFAPs while
the more complicated relationship between $g_0, h_0$ and $U$
for QSAPs breaks this link~\cite{Tailleur:2008:PRL}. This explains the
different roles of pressure in these two systems when considering change of ensembles.

\section{Conclusion}

In this article, we have shown how to derive the phase equilibria of MIPS for a number of different systems. At the hydrodynamic scale, the simple gradient terms that drive Active Model B~\cite{Wittkowski:2014:NC} out of equilibrium still allow for the construction of a generalized thermodynamics, which leads to the definition of generalized chemical potential, pressure and surface tension. Using this formalism, we account quantitatively for the binodal curve of fully-phase separated systems as well as for its finite-size corrections. 

For quorum-sensing active particles, we have shown how to build a hydrodynamic description that fits within our generalized thermodynamic framework, using a combination of a local mean-field approximation and a gradient expansion. Despite these approximations, our formalism   accounts quantitatively for the phase diagram of QSAPs. For particles interacting via repulsive pairwise forces, no closed hydrodynamics description including the relevant gradient terms exist in the literature. We thus followed an alternative route and showed how the binodals are selected by an equality of mechanical pressure complemented by a violation of the equilibrium Maxwell construction due to interfacial contributions.  

Our identification of the relevant intensive variables governing the phase equilibrium of MIPS is important to define thermodynamic ensembles, which we have illustrated by
considering the isobaric ensemble for QSAPs and PFAPs. We hope that
our approach will pave the way towards a more general definition of
intensive thermodynamic
parameters~\cite{Bertin:2006:PRL,Bertin:2007:PRE,Dickman:2016:NJP} for
active systems. Building a thermodynamic of active matter would
further improve our understanding and control of these intriguing
systems and has become a central question in the
field~\cite{Tailleur:2008:PRL,Palacci:2010:PRL,Brady:2014:PRL,Yang:2014:SM,Speck:2014:PRL,Wittkowski:2014:NC,Solon:2015:EPJST,Solon:2015:PRL,Solon:2015:NP,Ginot:2015:PRX,Takatori:2015:PRE,Bialke:2015:PRL,Farage:2015:PRE,Marconi:2015:SM,Fodor:2016:PRL,Dijkstra:2016:arxiv}.

{\em Acknowledgments:} APS and JS contributed equally to this
work. We thank M. Kardar and H. Touchette for discussions.  APS
acknowledges funding through a PLS fellowship from the Gordon and
Betty Moore foundation. JS is funded by a Project Grant from the
Swedish Research Council (2015-05449). JT \& AS ackownledge the hospitality of MLB Center for Theoretical Physics. MEC is funded by the Royal Society. This work was funded in part by EPSRC Grant EP/J007404.  YK is supported by an I-CORE Program of the Planning and Budgeting
Committee of the Israel Science Foundation and an Israel Science
Foundation grant. JT is funded by ANR Baccterns. JT \& YK acknowledge
support from a joint CNRS-MOST grant. The PFAP simulations were performed on resources provided by the Swedish National Infrastructure for Computing (SNIC) at LUNARC.

\appendix{}

\section{Hydrodynamics of QSAPs}
\label{app:drif-diffusion}

In this section we derive the hydrodynamic equations of QSAPs
interacting via a density-dependent velocity.  In the hydrodynamic
description, we consider only smooth density profiles, slowly varying
in space and time, so that we can expand the self-propulsion speed as
\begin{equation}\label{eq:vexp}
  v[\tilde\rho(\vec r_i+\eps \vec u_i)] \simeq v[\tilde \rho(\vec r_i)] +  \eps \vec u_i \cdot \grad v[\tilde \rho(\vec r_i)]
\end{equation}
Furthermore, for a system of size $L$, the (diffusive) relaxation time
$\tau_D$ of the density profile scales as $L^2$ and is much larger
than the microscopic orientational persistence time
$\tau=(\alpha+(d-1) D_r)^{-1}$. To construct the large-scale dynamics
of QSAPs, we first coarse-grain their dynamics on time scales such
that $\tau \ll t \ll \tau_D\sim L^2$, following the method detailed
in~\cite{Cates:2013:EPL,Solon:2015:EPJST}. In practice, we first
construct a diffusive approximation to the dynamics of QSAPs on a time
scale over which their density field does not relax so that the
propulsion velocity of a single particle depends on its position and
orientation through a function $v(\vec r_i,\vec u_i)$ which is
constant in time.

\subsection{Diffusion-drift approximation}
The probability $\psi(\vec r,\vec u)$ of finding a given particle at
position $\vec r$ with an orientation $\vec u$ evolves according to:
\begin{equation}
  \label{eq:master}
  \dot \psi=-\div\left[ v(\vec r,\vec u)\vec u\psi-D_t\grad\psi\right]+D_r\Delta_u\psi-\alpha\psi+\frac{\alpha}{\Omega}\int d\Omega'\psi
\end{equation}
$\psi$ can be expanded in spherical (3d) or Fourier (2d) harmonics:
\begin{equation}
  \label{eq:spherical-expansion}
  \psi(\vec r,\vec u)=\varphi+\vec p\cdot\vec u+ Q:M +\Theta[\psi]
\end{equation}
where $M_{ab}=u_au_b-\delta_{ab}/d$, and $\varphi$, $\vec p$, and $Q$
solely depend on $\vec r$.  $\Theta[\psi]$ is the projection of $\psi$
on higher order harmonics, which plays no role in the following. We
furthermore introduce the scalar product
\begin{equation}
  \label{eq:sclar-prod}
  \langle f,\,g\rangle =\int d\vec u f(\vec u)g(\vec u)
\end{equation}
where the integration is over the unit sphere. The components of
$\psi$ in the expansion~\eqref{eq:spherical-expansion} are then
obtained from
\begin{equation}
  \label{eq:orthogo}
  \langle 1,\,\psi\rangle =\Omega\varphi; \qquad \langle\vec u,\psi\rangle=\frac{\Omega}{d}\vec p; \qquad \langle M_{ab},\psi\rangle=\tilde\Omega Q_{ab}
\end{equation}
where $\Omega$ is the area of the unit sphere and
$\tilde\Omega\equiv \frac{2\Omega}{d(d+2)}$. Projecting
Eq.~(\ref{eq:master}) onto 1, $\vec u$ and $M$ yields the dynamics of
$\varphi$, $\vec p$ and $Q$:
\begin{eqnarray}
  \label{eq:phi}\dot\varphi&=&-\big\langle \frac 1 \Omega , \div v \vec u\psi\big\rangle +\div(D_t\grad \varphi)\\
  \label{eq:p}\dot{\vec p}&=&-\big\langle \frac{d \vec u}{\Omega}, \div v \vec u\psi \big\rangle
  +\div(D_t\grad \vec p)-\frac{\vec p}\tau\\
  \label{eq:Q}\dot Q&=&-\big\langle \frac{M}{\tilde\Omega},  \div v \vec u\psi\big\rangle+\div(D_t\grad Q)
  -\frac{Q}{\tau_Q}
\end{eqnarray}
where $\tau_Q^{-1}=2dD_r+\alpha$ is the relaxation time of the second
harmonic $Q$. Similar equations could also be derived for higher order
harmonics. However, the structure of Eqs.~(\ref{eq:phi}-\ref{eq:Q})
immediately shows that $\varphi$ is the sole hydrodynamic field since
all higher order harmonics relax on times of order ${\cal O}(1)$
($\tau$ for $\vec p$ and $\tau_Q$ for $Q$). Consequently, one can assume
that $\dot {\vec p}$ and $\dot Q$ vanish, as would the time derivative
of higher order harmonics. The structure of
Eqs.~(\ref{eq:p}-\ref{eq:Q}) then shows that all harmonics beyond
$\varphi$ are at least of order ${\cal O}(\grad)$ in the gradient
expansion.

Going further than Refs.~\cite{Cates:2013:EPL,Solon:2015:EPJST}, we
now also expand $v$ in spherical harmonics. Under~\eqref{eq:vexp},
only the first two harmonics matter and we use:
\begin{equation}
  \label{eq:dvlpt-v}
  v(\vec r,\vec u)\equiv v^0(\vec r)+\vec v^1(\vec r)\cdot \vec u
\end{equation}
where $\vec v^1$ will be of order ${\cal O}(\grad)$ in the gradient
expansion. Eq.~\eqref{eq:Q} then gives for $Q$:
\begin{eqnarray}
  \dot Q_{ab}&=&-\frac{d+2}{2}\left(\partial_c (v^0p_d+v^1_d\varphi)B_{abcd}+\partial_cv^1_dQ_{ef}C_{abcdef}\right)\nonumber\\
  &&\qquad+\partial_cD_t\partial_cQ_{ab}-\tau_Q^{-1}Q_{ab}+\partial_c\chi^Q_{abc}\label{eqQ}
\end{eqnarray}
where the constant tensors $B$ and $C$ are defined as
\begin{equation*}
  B_{abcd}\equiv\frac d \Omega \langle M_{ab},M_{cd}\rangle,\quad C_{abcdef}\equiv\frac{d}{\Omega}\langle M_{ab},u_cu_dM_{ef}\rangle
\end{equation*}
and $\chi_{abc}^Q\equiv \langle
M_{ab},u_cv\Theta[\psi]\rangle/\tilde\Omega$ stems from higher order
harmonics. Since $\dot Q \simeq 0$ on hydrodynamic time and space
scales, one finds at first order in gradients
\begin{equation}
  Q_{ab} \simeq O(\grad^2)
\end{equation} 
Similarly, the dynamics of $\dot p$ is given by
\begin{eqnarray}
  \dot p_a &=&-\partial_av^0\varphi-\frac{2}{d+2}\partial_bv^0Q_{ab}+\frac{1}{d+2}(\partial_bv^1_bp_a\\
&&\;+\partial_bv^1_ap_b+\partial_av^1_bp_b)+\partial_bD_t\partial_bp_a-\tau p_a+\partial_b\chi_{ab}^p\nonumber
\end{eqnarray}
where $\chi_{ab}^p=d\langle
u_a,u_bv\Theta[\psi]\rangle/\Omega$. Again, using $\dot {\vec p} \simeq
0$, one finds at first order in gradients
\begin{equation}
  \label{eq:p-grad}p_a=-\tau\partial_av^0\varphi+O(\grad^2)
\end{equation}
Finally, the dynamics of $\varphi$ reads
\begin{equation}
  \label{eq:phi-exp}\dot \varphi=-\frac{1}{d}\partial_a(v^0p_a+v^1_a\varphi)-\frac{2}{d(d+2)}\partial_av^1_bQ_{ab}+D_t\partial_a^2\varphi.
\end{equation}
Using Eq.~(\ref{eq:p-grad}) and $Q_{ab}=O(\grad^2)$, the dynamics of
$\varphi$ at diffusion-drift level reduces to the Fokker-Planck
equation
\begin{equation}
  \label{eq:drif-diff}
  \dot \varphi=-\div \left[\vec V\varphi-D\grad \varphi\right]
\end{equation}
with
\begin{equation}
  \label{eq:V-D}
  \vec V=\frac{\eps\grad v}{d}-\frac{\tau v \grad v}{d}; \qquad D=D_t+\frac{\tau v^2}{d}
\end{equation}

\subsection{Hydrodynamic equation}
The Fokker-Planck equation~\eqref{eq:drif-diff} for $\varphi$ is
equivalent to an It\=o-Langevin dynamics for the position of the
QSAP. From there, one can derive the collective dynamics of $N$ QSAPs
using It\=o calculus, as was done many times in simpler
settings~\cite{Dean:1996:JPA,Tailleur:2008:PRL,Solon:2015:EPJST}. For
simplicity, we consider here the case $D_t=0$. One then finds the
coarse-grained $N$-body density of QSAPs to follow the stochastic
dynamics
\begin{equation}
  \dot\rho=\div\Big[\rho D(\tilde \rho)\grad\big[\log \rho v(\tilde \rho)+\frac{\eps}{\tau v(\tilde \rho)}\big]+\sqrt{2\rho D(\tilde\rho)}\Gvec\eta\Big], \\
\end{equation}

We can now expand $\tilde \rho(\vec r)$ in gradients of the density field
\begin{equation}
  \tilde \rho \simeq \rho + \frac 1 2 \ell^2 \Delta \rho+{\cal O}(\grad^3)
\end{equation}
with $\ell^2=\int r^2 K(\vec r)d\vec r$. In turn, this implies for the
propulsion speed
\begin{equation}
  \label{eq:expansion-v2}
  v[\tilde\rho(\vec r)] \simeq v(\rho)+\ell^2v'(\rho)\Delta\rho+{\cal O}(\grad^3)
\end{equation}
Finally one finds the self-consistent dynamics for $\rho$:
\begin{equation}
  \dot\rho=\div\left[\rho D(\tilde \rho)\grad\left(\frac{\delta\cF}{\delta\rho}-\kappa(\rho)\Delta\rho \right)+\sqrt{2\rho D(\tilde\rho)}\Gvec\eta\right], \\
\end{equation}
where ${\cal F}[\rho]=\int d\vec r f[\rho(\vec r)]$ and
\begin{equation}
  \label{eq:f-kappa}
  f'(\rho)=\log [\rho v(\rho)]+\frac{\eps}{\tau v(\rho)}; \,\, \kappa(\rho)=-\ell^2\frac{v'(\rho)}{v(\rho)}\left(1-\frac{\eps}{\tau v(\rho)}\right)
\end{equation}
As hinted
before~\cite{Tailleur:2008:PRL,Wittkowski:2014:NC,Solon:2015:EPJST},
the non-locality of the density sampling results in a `surface tension
generating' term $\kappa(\rho)$. Interestingly, the asymmetric sensing
$\eps$ affects both the free energy density $f(\rho)$ and the gradient
terms $\kappa(\rho)$.

\section{PFAPs} 

\subsection{Constant-volume simulations}
\label{app:simu-cst-volume}

Simulations in the isochoric (constant-volume) ensemble were carried
out in rectangular boxes of size $L_x\times L_y$ with periodic
boundary conditions using a modified version of the LAMMPS molecular dynamics package\cite{Plimpton:1995:JCompPhys}. For simulations in slab geometry at coexistence, we chose
$L_x=500$, $L_y=300$,  and $N=$ 115,000 particles. In order to ensure a stable, flat (on average)
interface spanning the $\hat {\vec y}$-direction, these simulations were initiated
by first equilibrating the particles in a smaller box of size
$300 \times 300$ with $v_0 = 0$, yielding a near-close-packed
phase. After this initial equilibration, the box was expanded in the
$\hat {\vec x}$-direction and the activity was turned on, after which the system
relaxed towards a phase-separated steady state. The simulations were
run for a time $t=1000$, the data being collected over the second half of
this time.

We compute binodal densities by coarse-graining the local density using a weighting function
$w(r) \propto
\exp[-r^{2}_{\mathrm{cut}}/(r^{2}_{\mathrm{cut}}-r^{2})]$,
where $r$ is the distance between the particle and the measuring point, and
$r_{\mathrm{cut}}$ is a cut-off distance. Histograms of the density then
show two peaks that we identify as the coexisting densities.

In order to handle the relatively large fluctuations in the position of the interface, the density and pressure profiles measured on each timestep was translated to a common origin. This point was taken to be the point where the density (averaged over $y$) has fallen below $\rho = 0.95$. 

\subsection{Constant-pressure simulations}
\label{app:simu-cst-pressure}

Here, we used a simulation box with $L_y=100$, $N = 10000$ and periodic boundary
conditions in the vertical direction only. In the $x$-direction, the
system was confined by two walls modeled by harmonic potentials:
\begin{eqnarray}\label{U_wall}
  V_{w}^{R}(x) &=& k(x-x_R)^2 \Theta(x-x_R)\\
  V_{w}^{L}(x) &=& k(x-x_L)^2 \Theta(x_L-x)
\end{eqnarray}
where $k$ controls the stiffness of the walls (we take $k=5$). The
right wall is fixed at $x=x_R$, while the position $x_L$ of the
left wall obeys the deterministic overdamped dynamics:
\begin{equation}\label{v_wall}
  \dot{x}_L = \gamma \Big(P_w L_y -  \sum_{i=1}^N \frac{\partial V^\ell_w}{\partial x_i}\Big)
\end{equation}
where $x_i$ is the abscissa of particle $i$, $P_w$ the pressure
externally imposed on the wall and $\gamma$ its mobility, taken to be
$\gamma = 0.1$. The dynamics is integrated by Euler time stepping,
with the same time step as for the particles. To compute the phase
diagram in the isobaric ($N,P_w,\Pe$) ensemble, we ramp $P_w$ up and
down very slowly. For each $\Pe$, the system was first equilibrated at
a starting pressure $P_w$, which was then incremented or decremented
in steps of $\pm 1$ every $10^8$ time steps.

\subsection{Construction of the equation of state}
\label{app:EOS}

In order to separate the gradient contributions from the bulk
contributions to $P_A$ and $P_D$, one needs an accurate EOS for the full phase-separated parameter region -- something
which is not known \textit{a priori}. In order to obtain an
approximate EOS, we adopt a refined version of the strategy followed
in~\cite{Solon:2015:PRL}: we (\textit{i}) measure $P^0_A(\rho)$ and
$P^0_D(\rho)$ for $\mathrm{Pe} < \mathrm{Pe}_c$, where homogeneous
systems are stable for all densities, and (\textit{ii}) apply scaling
arguments to extend the validity of the EOS to $\mathrm{Pe} >
\mathrm{Pe}_c$.

We start from the exact expression of the active pressure in a
homogeneous system~\cite{Solon:2015:PRL}
\begin{equation}\label{P0_S}
\mu P^0_A(\rho) = \frac{v_0}{2D_r}v(\rho)\rho, 
\end{equation}
where $v(\rho)$ is the density-dependent single-particle swim velocity projected along its orientation~\cite{Solon:2015:PRL}
\begin{equation}
  v(\rho) \equiv v_0 +2 \mathds{I}_{xx}/\rho = v_{0} +\langle \vec{u}(\theta_i)\!\cdot\!\sum_{j\neq
    i}\vec{F}(\vec{r}_{j}-\vec{r}_{i})\rangle = \langle \dot \bfr_i \cdot \bfu_i \rangle\;.
\end{equation}
As has been shown several times
before~\cite{Fily:2012:PRL,Buttinoni:2013:PRL,Stenhammar:2013:PRL,Stenhammar:2014:SM,Solon:2015:PRL},
$v(\rho)$ is accurately described by a linearly decreasing function up
to near-close-packed densities. However, as the details of the
high-density region of the EOS are very important for the accuracy of
the predicted binodals, we furthermore include a quadratic term in
$\rho$ and a switching function which ensures a smooth transition to $v\rightarrow
0$:
\begin{equation}\label{v_rho}
v(\rho) = \frac{v_0}{2} (1 - s_1\rho + s_2\rho^2)(1-\tanh(s_3(\rho-s_4))),
\end{equation}
where $s_1 - s_4$ are fitting parameters which are found to be
essentially independent of $D_r$ (see Fig.~\ref{EOS}) as long as
we remain in the $\Pe < \Pe_c$ region. The local EOS for
$P_A^0$ is then given by~\eqref{P0_S} with~\eqref{v_rho}.

We now consider the local EOS for the direct pressure. For the values
of $v_0$ and $\epsilon$ used in this study, which control the
effective stiffness of the WCA potential, $P_D$ is found to be
independent of $D_r$ (see Fig.~\ref{EOS}). For \Pe = 40, we find that
$P^0_D(\rho)$ is accurately fitted by a biexponential function:
\begin{equation}\label{P0_D}
P^0_D(\rho) = d_1 \left(\exp(d_2 \rho)-1\right) + d_3 \left(\exp(d_4 \rho)-1\right),
\end{equation}
where $d_1 - d_4$ are fitting parameters. 

\vspace{20mm}
\bibliography{biblio-MIPS}
\end{document}